# Strong Purifying Selection at Synonymous Sites in *D. melanogaster*

David S. Lawrie[1,2], Philipp W. Messer[2], Ruth Hershberg[2,3], Dmitri A. Petrov[2]

[1]Department of Genetics, Stanford University, Stanford, California, United States of America
[2]Department of Biology, Stanford University, Stanford, California, United States of America
[3]Rachel & Menachem Mendelovitch Evolutionary Processes of Mutation & Natural Selection Research Laboratory, Department of Genetics, the Ruth and Bruce Rappaport Faculty of Medicine, Technion - Israel Institute of Technology, Haifa, Israel

Running head: Strong Selection at Synonymous Sites in Drosophila
Abbreviations: four-fold degenerate (4D)

Corresponding Author:

David Lawrie
Department of Genetics and Biology
Stanford University
371 Serra Mall
Stanford, CA 94305-5020
e-mail: dlawrie@stanford.edu
phone: (+1) (650) 736-2249
fax: (+1) (650) 723-6132

## Abstract

Synonymous sites are generally assumed to be subject to weak selective constraint. For this reason, they are often neglected as a possible source of important functional variation. We use site frequency spectra from deep population sequencing data to show that, contrary to this expectation, 22% of four-fold synonymous (4D) sites in *D. melanogaster* evolve under very strong selective constraint while few, if any, appear to be under weak constraint. Linking polymorphism with divergence data, we further find that the fraction of synonymous sites exposed to strong purifying selection is higher for those positions that show slower evolution on the Drosophila phylogeny. The function underlying the inferred strong constraint appears to be separate from splicing enhancers, nucleosome positioning, and the translational optimization generating canonical codon bias. The fraction of synonymous sites under strong constraint within a gene correlates well with gene expression, particularly in the mid-late embryo, pupae, and adult developmental stages. Genes enriched in strongly constrained synonymous sites tend to be particularly functionally important and are often involved in key developmental pathways. Given that the observed widespread constraint acting on synonymous sites is likely not limited to Drosophila, the role of synonymous sites in genetic disease and adaptation should be reevaluated.

## Introduction

As there are 64 codons and only 20 amino acids, most amino acids can be encoded by more than a single codon. Mutations that alter coding sequences (CDS), but do not alter amino acid sequences are referred to as synonymous mutations. Synonymous sites are then the collection of potential synonymous mutations present in a gene. Predicated on the assumption that the CDS of a gene is simply the recipe for making the protein, synonymous mutations were long thought to have no functional effect, in other words to be "silent" and thus selectively neutral [1, 2]. As a result, synonymous variation is often used as the neutral reference when measuring selection at functionally important, non-synonymous sites [3-7].

The observation of codon usage bias in many organisms was the first indication of possible functionality encoded by synonymous sites [8, 9]. Different codons for the same amino acid are often utilized at unequal frequencies across the genome. Highly expressed genes and codons encoding functionally important amino acids generally display particularly biased patterns of codon usage [9-11]. This observation led to the theory that selection for translation optimization generates higher levels of codon bias [12-15]. In other words, it is thought that the speed and accuracy of mRNA translation is higher for a subset of codons, referred to as "optimal" ("preferred") codons [14-19]. Such codons are translated more accurately and more efficiently because they are recognized by more abundant tRNA molecules with more specific anti-codon binding [14, 20, 21]. While this kind of selection acting on synonymous mutations is widely accepted, it is generally estimated to be weak - nearly, if not quite, neutral [22-31]. Synonymous variation is therefore still often thought to lack any major functional or evolutionary importance. In this paper, we further investigate the functionality of synonymous sites through detecting the action of purifying selection. If synonymous sites harbor highly deleterious variants under strong purifying selection, then that must change our view of the functional importance of synonymous sites and their potential role in genetic disease, as a possible source for adaptation, and as the neutral foil in tests for selection.

Previous tests for selection on synonymous sites have often been consistent with the presence of weak purifying selection operating on synonymous variation. Using the rate of divergence between species, the signal of purifying selection comes from a lower number of inferred substitutions on a phylogenetic tree at sites allowing for synonymous mutations, compared to the expectation provided by a neutral reference. Simply comparing the rate of evolution between a test and a neutral reference set can be problematic when weak purifying selection and mutational biases interact [32]. Nonetheless, synonymous sites do indeed appear to evolve slower than expected under neutrality for many organisms in a manner seemingly consistent with weak selection [22, 29-31, 33-39]. More evidence for weak selection acting at synonymous sites comes from the study of polymorphism within species. Purifying selection reduces the frequency of deleterious alleles in the population. To measure its effect, the site frequency spectrum (SFS) tabulates the fraction of observed SNPs in all frequency classes across the sites of interest. The overabundance of low frequency SNPs relative to the neutral expectation is the signal of purifying selection operating on the test sites. From this signal, one can calculate the strength of the selective force and the proportion of the test sites it affects [40, 41]. Such methods have been applied to studying the effects of selection on synonymous sites in a variety of Drosophila species, and have found evidence of weak selection - often favoring optimal codons [24-28].

Studies using divergence and polymorphism to infer selection as described above are, however, unable to detect the action of strong purifying selection. Tests that rely on divergence are limited in power to distinguish strong purifying selection from weak or moderate purifying selection. The problem lies in the efficacy of purifying selection (constraint) over a tree: a small, linear increase in the strength of constant purifying selection causes a large, exponential drop in the rate of evolution [42-44]. Weak to moderate constraint is thus capable of conserving sites over even large phylogenetic distances and increasing the number of species/tree length results in only a limited increase in power to distinguish strong from moderate or weak purifying selection [32]. Unlike tests on divergence that have difficulty distinguishing between strong and weak constraint, tests using the SFS of observed polymorphism can miss strong purifying selection entirely. While both weak and strong constraint eliminate variation from the population, strong selection does so far more efficiently. Therefore, at sites of strong constraint there are few SNPs and only at very low frequency in the population. Without sequencing enough members of a population to attain a deep sample, such SNPs will not be in the SFS of observed polymorphism. With no signal in the shape of the SFS from shallow population sequencing data, any strong selection acting on synonymous sites could not be detected via these methods.

While strong selection does not significantly affect the shape of a shallow-sample SFS, the lack of polymorphism can itself be a powerful signal of the action of selection [45, 46]. Knowing how many mutations should be present in the population sample, as compared to the amount actually present, can allow the estimation of the fraction of sites under strong selection. To do this, one needs a large sample of sites as the density of polymorphism is always low - on the order of a few percent. Differentiating between low densities in the test set and the neutral reference thus requires a large number of sites from each.

Note that both weak purifying selection and lower rates of mutation can likewise cause a paucity of SNPs. Ultra-low frequency variants can distinguish the signal of strong constraint from that of a variation in the mutation rate between the neutral reference and the set of sites being tested. While mutational cold spots only lead to a lower number of SNPs, under strong purifying selection some mutations should still be observable at very low frequencies in a deep enough sample of the population. Weak selection, meanwhile, will affect the shape of the spectrum beyond the rare alleles and can be estimated from that. Combined, the lack of polymorphism and the excess of rare variants from a genome-wide, deep sample, could give the necessary power to quantify the intensity of the strong constraint and the fraction of sites it affects. Thus, our dataset needs to include both a wide sample of sites from the genome, as well as a deep sample from the population.

The Drosophila Genetic Reference Panel (DGRP) for *D. melanogaster* provides such a dataset [47]. With 168 sequenced-inbred lines, this data set represents the whole genome (thus providing us with the widest possible sample of sites from the genome). The data also provides a deep sample of the variation within the *D. melanogaster* population of North Carolina. Using DGRP polymorphism, we estimate that, contrary to long held expectations, a substantial fraction of the synonymous sites in *D. melanogaster* is evolving under strong selective constraint. The discovery of strong selection on codon usage in Drosophila should dramatically change our collective perspective on the functional and evolutionary significance of synonymous sites.

## Results

To detect the action of selection on DGRP variation in synonymous sites, we need a neutral reference against which to compare the site-frequency spectrum and SNP density of synonymous sites. Short introns in Drosophila have been shown to be evolving neutrally or nearly so [48-51]. We therefore use sites from introns shorter than 86bp as the neutral reference, also removing the edges of these introns, 16bp away from the intron start and 6bp away from the intron end, as they may contain splicing elements [51].

For our collection of synonymous sites, to prevent any confusion of synonymous vs. non-synonymous selection acting on a given codon position, we focused on the third codon positions of the four-fold degenerate amino acids (Proline, Alanine, Threonine, Glycine, and Valine). All possible mutations in the third codon position are synonymous for these five amino acids. The third codon positions of these amino acids will from hereon be referred to as 4D sites. So that we could later relate our results from this analysis on polymorphism within *D.*

*melanogaster* to divergence across Drosophila, we used only those 4D sites from genes with 1-1 orthologs across the twelve sequenced Drosophila species as our test set [52].

To normalize the number of *D. melanogaster* strains sequenced at each position and any sequencing differences between short introns and 4D sites, we took only those positions which had their base pair called in at least 130 out of 168 strains and further resampled all SNPs to a depth of 130 homozygous strains (see Materials and Methods). The resulting data set consists of 863,972 4D sites, 5.58% of them containing a SNP, and 870,364 sites in short introns with 6.0% of these being polymorphic. By comparing the density and SFS of polymorphism between 4D and short intron sites, we can quantify the strength of selective forces operating on 4D sites and the fraction of such sites they affect.

Before doing so, several potential confounding factors to such an analysis need to be removed. The greatest of these is the difference in GC content between short introns and 4D synonymous sites. The GC content of 4D sites in *D. melanogaster* is 64%, compared to only 31% for short introns. Mutation is known to be generally biased towards A/T with particularly high rates of mutation from C:G to T:A [30, 50, 53-56]. With a higher GC content, 4D sites are thus expected to be subject to a higher mutation rate on average compared to short introns increasing their relative density of polymorphism. This mutational-GC effect could mask any effects of selection on 4D sites, which if present, would reduce the density of polymorphism in 4D sites compared to short introns.

A further complication is that there are spatial variations in the rates of mutation and recombination and in the amount and severity of linked selection across the genome [50, 57-60]. Sweeps, strong background selection, and variation in mutation rates may all influence the density of polymorphism in short intron sites relative to 4D sites [61, 62].

Outlined in Figure 1A is our bootstrap procedure to control for GC content and spatial variation in levels of polymorphism. We first pair 4D sites with short intron positions, requiring a short intron and 4D pair to have identical major alleles and be within 1KB of each other. Such pairs are then sampled with replacement, first drawing a 4D site and then picking at random one of its possible short intron partners, until the number of random pairs drawn equals the population of all 4D sites with short intron pairs. This process matches the GC content of the neutral reference to the test set, ensures the same spatial sampling of the intronic and 4D sites, and as a side bonus, normalizes the total number from each.

### Strong purifying selection on synonymous sites

Figure 1B shows the SFS for the SNPs in short introns and 4D sites from one bootstrap run. The shapes of the short intron and 4D spectra appear nearly identical. However, this similarity in the shapes of the spectra for 4D and short intron sites belies a large disparity in the density of polymorphism between the two sets. We measured the density of polymorphism in short intron and 4D sites and calculated the standard error of our measurement over 10 bootstrap runs. We find that 4D sites have approximately 22.1% (+/- 0.6%) fewer segregating sites as compared to short introns (Figure 1C).

To account for the relative paucity of polymorphism in 4D sites when the spectra of 4D and short introns SNPs are so similar, we combined both facets of information in a maximum-likelihood method allowing for the effects of multiple selective forces and demography on polymorphism (see Materials and Methods). We extended the SFS to include the number of non-polymorphic sites, the "zero"-frequency class, in our 4D and short intron bootstrap samples. Using such "amplitude" information along with the distribution of alleles over the observed frequency classes enables better maximum-likelihood estimation for parameters of strong selection. In this model, selection is parameterized by the effective selection coefficient $4N_es$: where s is the selection coefficient and $N_e$ is the effective population size of the organism. In our maximum-likelihood model, we used three categories of selection, neutral: $4N_es = 0$, weak purifying: $|4N_es| < 5$, strong purifying: $|4N_es| > 100$. The point estimates for the fraction of sites and the strength of constraint in each selection category can be seen in Table 1. While there is no evidence of extant weak selection acting differentially on 4D sites and short introns, ~22% of 4D sites are estimated to be under very strong constraint, $4N_es \sim -283$ +/- 28.3 (standard error estimate by bootstrap). When a coarse-grained demographic correction was applied to the SFS we obtained results that, though quantitatively are somewhat different ($4N_es \sim -370.1$ +/- 105), are qualitatively similar in that for both cases $100 \ll |4N_es| \ll 700$ – the calculable limit of our program (see Supplement – S1).

*Signal of strong selection not due to mutation*

One concern is the existence of some mutational difference between short introns and 4D sites beyond regional effects accounted for in the bootstrap, lowering the density of polymorphism in 4D sites relative to short introns. Such a difference is unlikely. As short introns are also transcribed, any transcription-coupled repair should affect both short introns and 4D sites equally, and its effects should be controlled for in our analysis. Furthermore, there have been no reports of such transcription-coupled repair lowering the mutation rate of transcribed regions in *D. melanogaster* [63]. The rate of mutation at a site may also be affected by that site's immediate neighbors, a phenomenon known as context-dependence [64-66]. However, when we used matching triplets in the bootstrap (i.e. the 4D site plus its immediate 3' and 5' neighbors paired against a similar trio of short intron sites), to account for any possible dinucleotide biases, we found no evidence of a mutational difference between 4D and short intron sites affecting our signal of strong constraint (see Supplement – S2.A).

Our maximum-likelihood estimation of the intensity of strong selection is itself evidence against a mutational force underlying the disparity in polymorphism between intronic and 4D sites. Simply lowering the mutation rate of 4D sites while maintaining their neutrality with respect to short introns would affect only the relative density of polymorphism and not the shape of the SFS. Therefore, a lower mutational rate at 4D positions acts equivalently to infinitely strong purifying selection (i.e. presence of lethals) in the above maximum-likelihood estimation. With our sample depth, the maximum-likelihood estimation has the power to detect the difference between a lower rate of mutation at 4D sites and strong constraint operating at $4N_es$ ~-300 (see Supplement – S2.B). Infinite selection or mutation on 4D sites would yield an estimate of $4N_es$ ~-700 (see Supplement – S2.B), far outside the range of any estimate of the strength of selection we made. The finite estimate of constraint that we obtained with and without demographic correction argues against the possibility that a mutational difference between 4D sites and short introns explains our results.

With no significant involvement of other forces eliminating synonymous polymorphism, the percent of missing variation in 4D sites is therefore a reasonable proxy for the fraction of sites under strong constraint. Thus ~22% of synonymous sites appear to be under very strong purifying selection in *D. melanogaster*.

**Evolutionary history of synonymous sites across Drosophila**

Exposing the action of the strong constraint on divergence between Drosophila species affirms the functional importance of these 4D sites across evolutionary history and reveals how these constrained synonymous sites evolve. If the strong constraint at 4D sites we identified within *D. melanogaster* has been constant across Drosophila, we would expect it to result in the complete conservation of the constrained 4D sites. If, on the other hand, the strong constraint is not constant and there is functional turnover at these sites, then we would expect to see substitutions occurring even at constrained sites along the Drosophila species tree. In order to compare the divergence between species to the constraint within a species, we considered only those 4D sites in amino acids conserved across the twelve Drosophila species from *D. melanogaster* to *D. grimshawi*. This simplifies the analysis as only the synonymous third position of the codon has been allowed to change over time. Thus, we can focus solely on the evolution of the synonymous site itself rather than consider the evolution of the entire codon. Figure 1C shows that the conservation of the amino acid has no bearing on the fraction of missing polymorphism in 4D sites. As such, the 4D sites of conserved amino acids provide a representative sample with which to study the strong constraint over the evolution of all 4D sites.

The gene orthologs in the other species were obtained from the 12 Drosophila Genome Consortium data realigned by PRANK [52, 67, 68]. We used the established 12 Drosophila species tree and re-estimated the branch lengths on the aligned 4D sites with PhyML (see Materials and Methods) [69]. From these alignments we removed the sequences belonging to *D. melanogaster* and *D. willistoni*. The *D. melanogaster* sequences were removed because the polymorphism data was extracted from this species and we wished to avoid a false concordance between the results from polymorphism and divergence. The *D. willistoni* sequences were removed, because the branch length leading to *D. willistoni* is long and the codon bias of *D. willistoni* is significantly different than from the rest of the twelve Drosophila species [70]. Having removed these species, the expected number of substitutions over the now ten Drosophila species tree for synonymous positions in otherwise conserved four-fold amino acids is estimated by PhyML at 3.1 subs/site [69]. To obtain site-wise estimates of conservation, we then inferred the number of substitutions along this tree for each 4D site independently using GERP (see Materials and Methods) [71, 72].

Figure 2 shows that the percentage of sites under strong constraint declines monotonically as the rate of evolution increases. 40.8% (+/- 1.9%) of completely conserved sites (0 substitution class), and only 7.1% (+/- 3.0%) of the fastest evolving sites (≥ 9.3 substitution class) are predicted to be under strong constraint. This difference in the amount of constraint between fast and slow-evolving sites allowed us to carry out a further control for any variation in mutation rate between short introns and 4D sites. We carried out an identical bootstrap procedure but pairing slow-evolving 4D sites with neighboring fast-evolving 4D sites instead of short introns as a neutral reference. We recapitulated our result of strong constraint at 4D sites by using slow- versus fast-evolving 4D sites (see Supplement – S2.C).

This correlation between a 4D site's conservation across species and strong constraint within a species underscores the functional importance of these synonymous positions over the evolutionary history of the Drosophila clade. However, over 80% of the sites currently under strong constraint in *D. melanogaster* fall outside the 0 substitution class, i.e. are not conserved across the ten Drosophila species. Indeed, over 11% of 4D sites under strong constraint in *D. melanogaster* have each acquired 6.2 or more substitutions over the tree, evolving quickly at more than twice the average rate. As even a moderate amount of selection results in complete conservation if it has been consistent over the tree, this suggests there has been functional turnover at these functionally important synonymous sites.

**Codon bias**

Codon bias is generally thought to be the product of background substitution biases combined with a weak selective force within genes skewing codon usage towards optimal (preferred) codons to increase translation efficiency and accuracy [19]. In Drosophila, translationally preferred codons are always G- or C-ending (except for in *D. willistoni*) [70]. The five four-fold degenerate amino acids have the following preferred codons: Alanine - GCC; Glycine - GGC; Proline - CCC; Threonine - ACC; and Valine - GTG [70]. Selection for codon bias is thus likely responsible for driving the GC content of 4D synonymous sites in *D. melanogaster* to 64% and to over 67% in the 4D sites of amino acids conserved over the 12 Drosophila species. While codon bias increases in conserved amino acids [17], as stated above, the strong selection at synonymous sites inferred in this paper does not (Figure 1C). To explore the relationship between codon bias and the strong constraint, we measured the fraction of sites under strong constraint within each codon, in unpreferred versus preferred codons conserved from *D. sechellia*-*D.grimshawi*, and across genes ranked by codon bias.

*The codon targets of strong selection*

Despite the fact that the conservation, and thus presumably the functional importance, of the amino acids appears not to matter, the fraction of 4D sites under strong constraint does fluctuate across the different amino acids: Alanine - 22.3% (+/- 0.9%); Glycine - 15.0% (+/- 1.8%); Proline - 18.0% (+/- 1.7%); Threonine - 24.8% (+/- 1.0%); Valine - 28.5% (+/- 1.2%). In order to identify the fraction of synonymous sites under constraint for an individual codon within each amino acid, we first assigned 4D sites to codons by their ancestral state, which we determined by parsimony using *D. sechellia* as the outgroup. A substitution between *D. sechellia* and *D. melanogaster* will cause a site to be unpolarizable. Because more monomorphic 4D sites than polymorphic ones are unpolarizable, simply removing unpolarizable sites would cause a shift in the density of SNPs in 4D sites and alter our signal of constraint. Therefore, these ancestrally ambiguous sites (< 8%) were assigned to their respective codons by their major allele so as not to remove sites during polarization.

Figure 3 shows the relationship between the amount of constraint in all the 4D sites for all codons grouped by their optimality and amino acid as well as the amount of each codon in the bootstrap analysis, reflecting the abundance of that codon in the genome. Two striking observations result from this analysis. First, while optimal codons are more frequently constrained than non-optimal codons, 25% (P) versus 18% (U) over all, for any individual amino acid the optimal codon may not have the highest fraction of sites under constraint. For Proline and Valine, the 4D sites of the unpreferred codons CCA/CCG (Proline) and GTT (Valine) are the most frequently strongly constrained. Second, there are some codons that have no apparent strong constraint on their 4D site - i.e. their 4D SNP density matches or exceeds the SNP density of short introns. These codons with seemingly neutral 4D sites are also used rarely in the genome relative to the other codons for that amino acid. These results are qualitatively similar when restricting the analysis to conserved amino acids (not shown).

There would thus appear to be strong selection on codon usage beyond the canonical selection for optimal codons. Figure 3 defines which codons are "favored" by strong constraint for each of the five four-fold amino acids. Sometimes these are also the previously defined optimal codons, but sometimes they are not. Even though there is propensity of strong constraint to affect particular codons, each four-fold amino acid has more than one codon with some fraction of its synonymous positions across the genome under strong constraint.

Our procedure polarizing sites by parsimony to a single species outgroup and then by major allele can misidentify the ancestral allele. Thus SNPs can be grouped with the wrong set of monomorphic sites, subtly changing the SNP densities across the codons. For instance, the negative fraction of sites under constraint - indicative of an excess of 4D polymorphism relative to short introns - for Proline's codon CCT is likely a product of this mispolarization. It is more likely that codon CCT is similar to GGG, GGT, ACT, and GTA and has a neutral or nearly neutral level of polymorphism. Thus while the relationship between codons is worth noting, the exact numerical fraction of sites under constraint for each individual codon are all slightly biased beyond the nominal standard error. This bias is difficult to quantify but is not expected to be strong for most codon categories as *D. sechellia* and *D. melanogaster* are closely related species with few substitutions to throw off polarization. To eliminate any such biases from mispolarization and concurrently study the long-term signals of selection on 4D sites with respect to codon optimality and strong constraint, we refocused our analysis on only conserved codons.

*Conserved codons: The dueling signals of codon bias and strong constraint*

To more accurately quantify the relationship between selection for optimal codons and strong constraint, we restricted our bootstrap to include only those 4D sites from conserved amino acids in the 0 substitution class - i.e. those 4D sites conserved across the ten Drosophila species from *D. sechellia* to *D. grimshawi* (excluding *D. melanogaster* and *D. willistoni*). In such conserved codons, there are only a few substitutions along the *D. melanogaster* lineage at the 4D sites. In over 98% of these conserved 4D sites, both segregating and monomorphic, *D. melanogaster* shares an allele with the ten Drosophila species outgroup - the ancestral allele by parsimony. Such support as ten species sharing the same allele provides more confidence in the polarization of the SNPs at these sites. Also, roughly the same percent of monomorphic as polymorphic sites are removed as unpolarizable (less than 2% each) so that the act of polarization itself does not affect the density of 4D SNPs. This restriction allows for confident polarization by parsimony without changing the relative density of SNPs between 4D sites and short introns. As there are too few such conserved 4D sites to analyze each codon individually as above, we only consider the broad classes of preferred and unpreferred 4D sites. Note, however, these fully conserved codons are precisely where the action of selective forces has been most efficacious over evolutionary history.

In contrast to the results from all codons, when limiting the analysis to conserved codons (Figure 4), a higher fraction of unpreferred than preferred 4D sites are under strong constraint - 53% (U) to 38% (P). However, 4D sites in the optimal state in *D. melanogaster* have been conserved across the ten Drosophila species to a greater extent, almost three times as often, than their non-optimal counterparts (Figure 4). This is expected because weak selection for codon bias in other Drosophila species on the Drosophila tree is expected to generate conservation at optimal codons over and above the strong constraint we identify in this paper. Therefore, although fewer non-optimal codons are conserved in total, more of the conserved non-optimal codons have been so conserved because of the strong constraint.

*Strong constraint across genes ranked by codon bias*

To compare the gene targets of selection for codon bias and the strong constraint, we ranked genes by their Effective Number of Codons (ENC) [11] and Frequency of Optimal Codons (FOP) [20], each obtained from the database SEBIDA [73]. While neither metric accounts for local GC content, we used them to broadly classify genes by the extent of their codon bias (high, medium, low). We then performed 10 bootstrap runs on all the 4D sites within each gene-class. From Table 2, we can see that highly biased genes (having a high FOP and low-med ENC) have a slightly lower fraction of sites under strong constraint than genes with lower codon bias. Thus strong constraint acting on synonymous sites in *D. melanogaster* operates largely independently from canonical codon bias.

**Strong constraint as a function of different genic features**

Table 3 summarizes our analyses of how the extent of strong constraint is influenced by different genic features such as gene length, the location of the synonymous site along the gene, the chromosome on which the gene is located, whether or not the synonymous site falls within splice junctions, and nucleosome binding. Many of the associations below, while suggestive, are marginal in effect. The dominant pattern is that strong constraint at synonymous sites appears to be ubiquitous across different gene classes and functional elements within genes.

*Spatial distribution of strong constraint within genes*

Looking at the distribution of constrained sites within genes, we focused on those sites that are within 75bp from the start or stop codon and compared them to the 4D sites that lie in the middle of the gene. For this comparison, we took only those genes with a CDS longer than 150bp. ~31% of 4D sites near the translation start and stop are under strong constraint. This is nearly a 50% increase in the fraction of sites under strong constraint as compared to the middle of the gene where only ~21% of 4D sites are under strong constraint on average. Breaking the spatial distribution of 4D sites across the middle of the gene into finer segments, we find no other peaks of strong constraint beyond those at the 5' and 3' edges of the genes (see Supplement – S3).

*Bulk Nucleosomes*

Bulk nucleosomes wind themselves over ~146bp of DNA, attaching at semi-regular intervals. Their attachment points are associated with both lower mutation rates and selection [74-78]. Canvassing all 4D sites in the 146bp regions around known bulk nucleosomal binding sites [79], we find a small increase in the fraction of missing polymorphism in these 4D sites bound by nucleosomes (Table 3). However, we have reason to believe that this slight increase above 22% is due to weak selection acting on nucleosome-bound sites and is likely not related to the strong constraint we measure in this paper (see Supplement – S4). This potential weak-selective force does not impact our other results as it affects both short introns and 4D sites and we only measure selective differences between short introns and 4D sites.

*Splice Junctions*

To investigate whether strong constraint can be explained by the need to maintain splice junctions, we tested 4D sites near intron-exon splice junctions - i.e. within 48bp of a splice site. Around 26.0% of such 4D sites are under strong constraint (Table 3). This might indicate a role for splicing enhancers in the strong constraint, but Table 3 also shows that multi-exon genes and single-exon genes have similar amounts of strong constraint. The inference on the single-exon genes is particularly noisy, especially so given that our bootstrap method controls for distance to short introns. However, only about one-fifth of our 4D sites fall near splice sites and the modest enrichment of constraint near splice sites is not enough to explain the ubiquitous constraint at 4D sites across the genome or especially in single-exon genes.

*Gene length*

Longer genes tend to have slightly more sites under strong constraint than shorter genes (Table 3). Interestingly this correlation is stronger when taking intron and UTR length into account than when considering the CDS sequence alone. This pattern is the opposite of what is seen for codon bias in Drosophila [15, 80].

*X-linked vs. autosomal genes*

In Table 3, we show that X-linked genes have a lower fraction of sites under strong constraint than autosomal genes. This pattern is again the opposite of what is seen for codon bias [28, 81]. As selection is more efficient on the X chromosome [82], the cause for this difference is not clear and might reflect some difference in the types of genes located on the X as opposed to the autosomes.

**Strong constraint correlates with gene expression level over development**

To map how strong constraint at synonymous sites varies with gene expression over development, we ranked genes by their expression levels at each developmental time point in the ModEncode data set [83]. We split the genes evenly into three categories of expression - highly, moderately, and lowly expressed - within each developmental stage and ran 10 bootstraps for the 4D sites of the genes within each expression category in each developmental time point. The results are shown in Figure 5.

The overall gene expression level across development correlates well with the fraction of sites under strong constraint with lowly expressed genes tending to have fewer sites under strong constraint and highly expressed genes tending to have more sites under strong constraint. This pattern is strongest for the genes expressed highly in mid-late embryos, pupae, and adult males. The association of strong constraint with these developmental stages is further enhanced when the “high” expression group has been split in half into “high” and “very high” expression level categories (see Supplement – S5). In contrast to this preference of strong selection for genes highly expressed in embryo, pupal, and adult stages, codon bias is highest for genes whose expression peaks in larval stages [84].

**Strong constraint over gene ontology**

The difference in density of polymorphism between 4D sites and short introns does not allow for precise measurements of constraint on the synonymous sites of single genes. To identify a set of genes that are under particularly strong constraint at synonymous sites, we ranked genes by the fraction of their conserved amino acids that are unpreferred and conserved from *D. sechellia to D. grimshawi*, in the 0-substitution class (see Materials and Methods). Our method left 4,877 genes capable of being ranked of which we took the top sixth (812 genes, see Supplement – S7) as our gene set enriched for strong constraint.

To validate our method of selecting genes under strong constraint, we checked that our 812-gene set is indeed enriched for strong constraint at 4D sites. We performed a bootstrap analysis on the 4D sites of variable amino acids in the genes in and out of this top set. Estimating constraint using 4D sites from variable amino acids provides a measure of the fraction of synonymous sites under constraint independent from our surrogate using conserved amino acids. In the top 812 genes, we find a ~30% reduction in polymorphism at 4D sites in variable amino acids; in all 4,065 genes not in the top 812 set, we find an average of ~21% of 4D sites in variable amino acids under strong constraint. As such, our top 812 genes are enriched for almost 50% more 4D sites under strong constraint than the average gene. Note that any individual gene in the 812-set does not necessarily have elevated levels of strong constraint at its synonymous sites, nor does any individual gene of the 4,065 necessarily have a lower fraction of 4D sites under strong constraint.

In order to examine whether genes under strong constraint at synonymous sites tend to be enriched for certain functions, we used DAVID 6.7 [85, 86]. DAVID takes all the genes in the background data set (4,877 genes) and all genes in the test data set (812 genes) and looks for the enrichment of biological terms and gene families in the test set relative to the background. In Table 4, we list a subset of those biological terms found by DAVID's functional annotation clustering run on high stringency (for full information on the top 13 clusters, see Supplement – S6). We find that in genes enriched for strong constraint, we co-enrich for many important functional gene sets. In particular, we co-enrich for genes critical in pupae-to-adult morphogenesis and in late embryogenesis. This finding is consistent with the result that genes expressed highly in late embryos, pupae, and adults have elevated levels of strong constraint at 4D sites. Many other functional classes important to the basic development and functioning of *D.*

*melanogaster* appear to have a higher fraction of synonymous sites under strong constraint including: transcription factors, ribosomal genes, immunoglobulin genes, genes regulating gamete production – particularly oogenesis, cell-signaling genes – particularly synaptic transmission, and more.

## Discussion

The strong constraint at synonymous sites in *D. melanogaster* measured in this paper represents a powerful force. We estimate that ~22% of synonymous sites are experiencing, on average, a selective pressure between $4N_es$ ~ -250 – -500 against deleterious mutations. This strength of selection is as strong or stronger as has been measured via population genetic techniques at any class of sites, including non-synonymous ones [45, 46]. Mutations at strongly constrained synonymous sites should never rise above low frequency in the population and certainly will never fix, barring tight linkage to a very advantageous allele or a shift in the functional properties of the site. While detectable within a population, these mutations are effectively lethal over evolutionary time.

We tested a number of controls to rule out the possibility that our observation of strong purifying selection results from other forces with possibly similar signals: A lower mutation rate, for example, can cause a signal indistinguishable from strong selection in polymorphism if the sample depth of the population is too shallow. To account for this, and at the same time account for any variation in the amount of linked selection between 4D sites and short intron sites, we used a bootstrap to control for GC content and distance between the 4D and short intron sites. We also performed bootstraps controlling for dinucleotide content between 4D and short intron sites and performed bootstraps pairing slow-evolving 4D sites against fast-evolving 4D sites as the neutral reference. Neither revealed a mutational force underlying the ~22% drop in 4D polymorphism compared to short introns. As revealed by simulations, the finite estimate we obtained of the strength of strong selection is itself evidence against a mutational force being responsible for our signal, as mutational variation would behave like infinite selection on 4D sites. While we do not have the frequency depth from the population necessary to estimate a full distribution of selection coefficients for the strong constraint force, our point estimate of $4N_es$ ~-283 for these 22% of sites is statistically significantly different from the value of $4N_es$ ~-700 (the computational limit of our program) expected if the signal was due to variation in mutation rate.

We also controlled for deviations from mutation-selection equilibrium affecting both the 4D and short intron site frequency spectra using a frequency-dependent correction. Such deviations include demography, shared (linked) selection between 4D sites and short introns, and our own approximations to the SFS. Controlling for these deviations resulted in a higher estimate of the strength of selection ($4N_es$ ~-370) with larger error bars, but still significantly far from the boundary of $4N_es$ ~-700.

A constant influx of weakly advantageous alleles in coding sequences, as is expected to occur in *D. melanogaster* [59], could affect variation at nearby 4D sites more than at short introns. The resulting genetic draft generated by adaptive substitutions in coding sequences would weaken the apparent intensity of purifying selection on 4D sites by bringing strongly deleterious alleles to higher frequencies, making our above estimates of selection intensity conservative [87]. Even so, strong selection, rather than a mutational difference, would still underlie our signal, as genetic draft cannot alter the frequency of synonymous mutations that are simply absent from the population. On the other hand, sweeps of weakly advantageous alleles in coding sequences could eliminate polymorphism in 4D sites more so than in short introns. Narrow selective sweeps in coding sequences reducing variation at otherwise neutral 4D sites is, however, an unlikely explanation for our observations. When comparing 4D sites from different substitution rate classes against each other, we found a signal of strong constraint at conserved 4D sites relative to fast-evolving 4D sites. As sweeps should not affect the overall substitution rate of linked sites, strong purifying selection on synonymous sites is the best explanation for the lack of polymorphism at 4D sites relative to short introns.

Our ability to detect strong selection and differentiate it from other forces critically depends on the availability of deep and genome-wide population data. Previous data sets could only find weak or no constraint, thus always confirming our collective biological intuition that synonymous sites had little functional or evolutionary importance. In a shallower sample of even genome-wide data, the highly deleterious variants would be simply missing from the sample and there would be no power to distinguish strong selection from a variation in the rate of mutation. As an example, we simulated 4D sites evolving under the selective regime inferred from the real data (22% of sites at $4N_es = -283$) but with only 60 instead of 130 homozygous strains. Attempting to re-estimate the strength of selection from such a shallow sample results in the observation of seemingly infinite selection operating on 22% of 4D sites. Simulating 60 strains under a scenario where neutral 4D sites have a 22% lower mutation rate than do short introns results in the same inference of infinite selection. Genome-wide, deep population data sets were not available before recently and thus strong constraint could never before be unambiguously detected at synonymous sites.

Interestingly, the strong constraint in *D. melanogaster* appears to be a largely orthogonal force to codon usage bias, favoring an overlapping, but different set of codons with subtly different gene targets. Codon bias increases as the conservation of amino acids increases, while the strong constraint targets the 4D sites of both conserved and variable amino acids equally. We further identified the codons under strong constraint and, for any given amino acid, the codon(s) with the highest fraction of sites under constraint were not necessarily the optimal codon. Other studies have likewise noted signals of selection favoring non-optimal codons in Drosophila [25, 30, 33, 88, 89]. Overall, preferred 4D sites do have greater amounts of strong constraint acting on them, but the strong selective force targets a substantial fraction of the unpreferred 4D sites as well. There is also a weak anti-correlation between genes with a high fraction of constraint and genes with high codon bias, which extends to various gene features. Long genes are associated with higher levels of strong constraint at 4D sites as opposed to shorter genes, in opposition to codon bias in Drosophila [15, 80]. X-linked genes have a lower fraction of 4D sites under constraint than autosomal genes, wheras codon usage bias is stronger on the X [28, 81]. While both codon bias and the fraction of 4D sites under strong constraint are correlated with highly expressed genes, codon usage bias is strongest in genes with their highest expression in larval stages [84] as opposed to the strong constraint seen most often in genes expressed highly in mid-late embryo, pupal, and male adult stages.

The pattern of conservation over 4D sites supports the existence of canonical weak selection in Drosophila favoring the conservation of preferred 4D sites across the twelve species, but it appears to have been relaxed in *D. melanogaster*. In our SFS analysis, we were not only able to gauge the intensity of strong selection, but also show a lack of contribution from weak purifying selection to our signal. If any weak selection is still acting differentially on synonymous sites relative to short introns, then the signal is not powerful enough to be detected by our SFS model or contribute much to our signal of lost polymorphism. These results recapitulate some earlier results on *D. melanogaster* [24], although see [25-28]. While weak selection on 4D sites in *D. melanogaster* may not have vanished completely, the large influx of mutations and substitutions away from optimal codons corroborates some relaxation of constraint for codon bias in *D. melanogaster* [25, 30, 31, 33, 38]. Overall, weak selection for codon bias would seem to be less of a force on synonymous sites in *D. melanogaster* than in its sister species where weak selection for codon bias can be detected with far less ambiguity [24, 30, 31, 33, 34]. Thus, evidence suggests that there are at least two major, orthogonal forces affecting the evolution of 4D sites in Drosophila: the weak selective force driving codon bias that favors optimal codons, present in other Drosophila species, but relaxed in *D. melanogaster*; and an extant strong selective force targeting both optimal and non-optimal codons in *D. melanogaster* and across the Drosophila phylogeny.

The presence of splicing enhancers and nucleosomes do not explain the pattern of strong purifying selection. However, the function underlying the strong constraint of synonymous sites may yet prove to be acting at the level of gene regulation. Those genes where strong selection on synonymous sites acts most frequently are often highly expressed regulatory proteins, operating in essential, tightly controlled developmental pathways. These are genes where the regulation of gene expression will matter most. Regulation of gene expression may be acting at the level of mRNA structures, mRNA stability, miRNA binding sites, and the modulation of translation rate [90-98]. Choice of synonymous codons might affect all of these levels of gene regulation. It should be noted that these various hypotheses are not mutually exclusive and may be intertwined. mRNA structures - as well as their avoidance - may be involved in translation initiation, modulation of mRNA half-life, and accessibility of the mRNA to proteins and miRNAs [97-99]. miRNAs themselves have a host of different functional effects in different species and different genes within a species but are well known in their role of mRNA degradation [100, 101]. Dynamics of translation not only affect the overall rate at which proteins are created, but also affect how these proteins fold and even the mRNA half-life [90-93, 102-104]. The possibility that strong selection acts at the level of modulating translation rate through the presence of slow/fast sites is interesting as the translation speed of a codon is not necessarily related to codon optimality [95, 105, 106]. Given the pattern of the strong constraint across the different codons both optimal and non-optimal, the strong selective force may be due to the abundance of wobble vs. Watson-Crick tRNAs available for that codon. Ascertaining the functional mechanism underlying the observed strong constraint acting on synonymous sites could reveal deep insights into the regulation of gene expression.

Regardless of the specific functional mechanism underlying the strong constraint, experimental evidence from a wide range of species substantiates an important functional role for synonymous sites. Directed mutagenesis studies targeting synonymous sites as well as studies of natural polymorphism have found consequential changes in protein levels and functionality due to natural synonymous variation and induced mutations [103-105, 107-117]. In an experiment done on the Alcohol dehydrogenase (Adh) gene in *D. melanogaster*, changing 10 wild-type preferred Leucine alleles to unpreferred alleles in the 5’ region of the gene lowers the enzymatic activity of collected Adh by 25% [111]. The authors proposed that disruption of the sites’ translational efficiency and accuracy caused the drop in activity, but also noted that the functional effect was far larger than expected given the assumption of only weak selection on synonymous sites [111]. ‘Humanized’ versions of protein coding sequences, with codons replaced with synonymous, putatively optimal codons in humans, show much greater protein expression and function when transfected into mammalian cells than the originals or synthetic versions using a non-mammalian species' set of optimal codons [107-110]. Human gene *Multidrug Resistance* 1 (*MDR*1) contributes to the drug resistance of cancer cells [114]. Both naturally occurring alleles as well as induced novel mutations at synonymous sites in *MDR*1 affect the resulting protein's conformation, altering its substrate specificity in human cell lines [114]. In the *E. coli* gene *ompA*, exchanging eight frequently-used codons for synonymous infrequently-used codons near the gene start results in a 3-fold reduction in mRNA levels and a 10-fold reduction in synthesis of protein OmpA [104]. Meanwhile exchanging codons with low-abundance tRNAs to synonymous codons with high-abundance tRNAs in *E. coli* gene *sufI* - or increasing the abundance of those tRNAs - results in misfolding of the protein *in vitro* and *in vivo* [102].

What about the presence of strong constraint in the synonymous sites of other species? In addition to the above functional assays, there are reported to be a significant fraction of synonymous sites under an unknown intensity of constraint in many species [22, 29, 35-37, 39, 96, 118] and there is evidence for strong selection in humans [46]. For example, when compared to "neutral" controls, there is a reduction in polymorphism density and/or a lower rate of divergence at synonymous sites for many tetrapods including chicken, hominids, murids, and mammals in general [22, 29, 35-37, 118]. Further, some of these species have undetectable or weak levels of codon bias, presumably commensurate with their small effective population sizes and thus the weakness of selection in favor of optimal codons [36, 119]. Using a similar model to the one described in this paper, Keightley and Halligan (2011) found evidence to support that weak selection alone is unable to explain the pattern of diversity at 4D synonymous sites in humans [46]. While that study lacked the sample depth of polymorphism to be able to gauge the intensity of the strong selection, they estimated that 11% of 4D sites are evolving under a strong selection regime of $4N_es > 40$ [46]. Our results from Drosophila with a deeper population sample lend credence to the hypothesis that, in humans too, a force of strong constraint is responsible for the lack of polymorphism at 4D sites rather than a mutational force or other confounding factors. For many species, there has been no conclusion that the constraint on their respective synonymous sites is strong, but many of the signals are consistent with what we find in Drosophila with the fraction of sites under constraint, the amount of missing polymorphism, and the lack of relationship to codon bias. Thus with genome-wide, deep population SNP data becoming available for many of these other species, we may well find strong selection on synonymous sites to be ubiquitous.

As synonymous sites have often been used as the neutral reference in tests for purifying and adaptive selection, many estimates of the fraction of sites under constraint in other classes, such as non-synonymous sites, UTRs, and many others, are likely to be conservative. This result from population genetics supports findings that synonymous sites may harbor many, important causal variants and that studies ignoring the potential contribution of synonymous mutations may be likewise unnecessarily conservative [90]. Turnover at these strongly constrained synonymous sites could also represent a significant source of interspecies functional divergence and adaptation. The potential of synonymous sites to be sources of adaptation and genetic disease merits further investigation. Although the functionality underlying this strong constraint remains unknown, recent studies have uncovered a myriad of different types of functional information encoded into the CDS of genes beyond the protein recipe, including controls for translational efficiency and accuracy, splicing enhancers, micro-RNA binding, nucleosome positioning, and more. With the discovery of a significant fraction of sites under strong constraint in Drosophila, two things become clear: the role of synonymous sites in the biology of genomes is far greater than the neutral, "silent" part they were once assumed to play; and we still have much to learn about the functionality encoded in genes.

## Materials and Methods

### Data

The SNP data set from DGRP (http://dgrp.gnets.ncsu.edu/data/) consists of 168 inbred lines from a population of North Carolina *D. melanogaster* [47]. The SNPs were annotated as synonymous, non-synonymous, and intronic using Flybase release 5.33 (ftp://ftp.flybase.net/genomes/Drosophila_melanogaster/dmel_r5.33_FB2011_01/) [120]. If a position was found in multiple gene annotations, only those sites where the SNP was synonymous in all sites was called synonymous. Short intron sites are defined as those sites falling in introns of less than length 86bp, 16bp away from the intron start and 6bp away from the intron end in order to eliminate any functional sequences at the edges of the introns [51]. Eliminating 16bp from each side did not change SNP density (not shown). Any remaining purifying selection, especially strong purifying selection, in short introns makes our results more conservative. Four-fold (4D) sites are the collection of 3rd codon positions for the following amino acids: Proline, Alanine, Threonine, Glycine, and Valine.

All sites were resampled to a depth of 130 strains. All sites with sequence information for fewer than 130 strains were excluded. For SNPs at sites with more than 130 strains or which contained heterozygous lines at that position, a 130 allele subset was chosen randomly. If the SNP was no longer polymorphic after this random resampling, that position was moved into the non-polymorphic site class. We also removed any position with more than 2 alleles present.

We restricted our analysis to genes with 1-1 orthologs across the 12 Drosophila species tree [52] and where the longest transcript annotation had remained intact in release 5.33 - even if it is no longer the longest transcript in release 5.33. We used the remaining 5,709 coding sequences aligned with PRANK from Markova-Raina and Petrov (2011) [67, 68].

**SFS – maximum likelihood and simulation**

To determine the distribution of selective effects on a group of sites based on the shape and the amplitude of the SFS, we assume a two-state framework where sites are either monomorphic in the wild-type state or polymorphic with a neutral or deleterious mutation at some observed frequency in the population. Using short introns as a neutral reference, our model aims to capture the fraction of synonymous sites falling into three broad selection categories – those with neutral, weakly deleterious, or strongly deleterious mutations – and estimate the effective selection coefficients acting on those mutations.

Strong constraint can be difficult to capture as strong selection has a greater effect on the amplitude of the SFS, the total number of observed mutations, than on its shape, the frequency distribution of observed mutations. Using a similar expansion to the standard SFS to Keightly and Eyre-Walker (2007) [45], we add the zero-frequency class, the fraction of monomorphic sites, to the SFS. The SNP density provides the additional information necessary to infer the action of strong constraint.

Equal to $4N_e\mu$, $\theta$ is mutation rate scaled by the effective population size and determines the neutral SNP density. The short intron SFS, used as neutral reference, anchors our estimate of $\theta$ which in turn allows us to estimate the amount of missing synonymous polymorphism in each selection category, *c*. As purifying selection increases, the overall density of observed polymorphism is reduced in the fraction of 4D sites in that selection class and the expected distribution of mutation is further skewed towards rare frequencies in the population. Each category has a single selection parameter, $\gamma_c$, a point estimate of the effective strength of selection, $4N_es$, operating on the 4D sites in that class. For those 4D sites in the neutral category, $\gamma_c = 0$. For those in the weakly deleterious category, $0 < |\gamma_c| < 5$. For those in the strongly deleterious category, $|\gamma_c| > 5/100$ – the choice of boundary did not affect results.

For our sample of *n* chromosomes from the population, assuming mutation-selection balance, we have the following analytical prediction for the SFS, *g(x)* – the expected fraction of 4D sites with SNPs at frequency *x* in the sample [42]:

(1) $$g(x,c) = \theta \bullet f_c \bullet L \frac{\left(1 - e^{-\gamma_c \bullet (1-x)}\right)}{\left(1 - e^{-\gamma_c}\right) x(1-x)}$$

(2) $$if\ \gamma_c = 0\ then\ g(x,c) = \theta \bullet f_c \bullet L/x$$

*g(x,c)* is the contribution of each selection category to the overall SFS. *L* is the total number of 4D sites while $f_c$ is the fraction of 4D sites in each selection category *c*.

(3) $$g(x) = \sum_c g(x,c)$$

(4) $$if\ x = 0\ then\ g(0) = L - m\ where\ m = \sum_{x=1/n}^{x<1} g(x)$$

*g(0)* are the zero-frequency class, monomorphic, sites and are what gives the SFS "amplitude" information – the density, rather than just the shape, of the spectrum. While *m* is the total number observed SNPs in the sample.

The theoretical SFS for intronic sites is the same as above, only all sites are assumed to be neutral. However, any real SFS does not reflect the true frequency distribution of the SNPs in the population, but rather a binomial sampling of those SNPs and frequencies. The above is thus an approximation, as the probability of a site with a SNP at a given frequency in the sample from the population is not quite the same as the probability of a site with a SNP at a given frequency in the population as a whole. However, it is much more computational efficient for both speed and memory to use the approximation.

With this theoretical prediction of the distribution of sites over each frequency class in both the neutral reference (short intron SFS) and test set of sites (4D SFS), we can use maximum-likelihood to fit the parameters of our model to real data sets. Our model has 5 free parameters: $\theta$, ($\gamma_{weak}$, $\gamma_{strong}$), and ($f_{neutral}$, $f_{weak}$, $f_{strong}$) where $f_{neutral}$ = 1-$f_{weak}$-$f_{strong}$. The total likelihood, $\lambda$, of the model's fit to the data, *D*, is equal to product of the fit the short intron and 4D sites spectra:

(5) $$\lambda_{full}\left(D \mid \theta, \bar{\gamma}, \bar{f}\right) = \lambda_{4D}\left(D \mid \theta, \bar{\gamma}, \bar{f}\right) \times \lambda_{SI}\left(D \mid \theta\right)$$

$\lambda_{4D}$ and $\lambda_{SI}$ are the likelihood of the observed SFS given the expected SFS as determined by the free parameters and equations (1) – (4). These likelihoods are the multinomial probability of observing a certain number of sites, *k*, with SNPs in frequency class *x* in the sample given theoretical expectations. Taking short intron sites as an example (same for both):

(6) $$\lambda_{SI}\left(D \mid \theta\right) = \prod_{x=0}^{x=\frac{1}{2}} \left(p(x \mid \theta)\right)^{k_x} \ where\ p(x \mid \theta) = \begin{cases} g(0)/L & if\ x = 0 \\ g(\frac{1}{2})/L & if\ x = \frac{1}{2} \\ \left(g(x) + g(1-x)\right)/L & o.w. \end{cases}$$

Equation (6) is thus the probability that the folded theoretical SFS, *g(x)*, matches the empirical folded SFS, $k_x$. We folded the spectrum to avoid any problems with inferring the ancestral state.

We then maximized the parameters $\theta$, ($f_{neutral}, f_{weak}, f_{strong}$), and ($\gamma_{weak}, \gamma_{strong}$) in Matlab using fminsearch, an implementation of the Nelder-Mead simplex method [121], on the negative log-likelihood of $\lambda_{full}$. The observed spectra were obtained from the bootstrapped 4D and short intron pairs. Where simulations were needed in this study, theoretical spectra were calculated using the above equations (1)-(4) and then the parameters were re-estimated by the outlined maximum-likelihood procedure on those theoretical spectra acting in place of the empirical data.

*Frequency-dependent correction of SFS*

We also employed a frequency-correction developed in Eyre-Walker et al (2006) [40] to control for demography or any weak and linked selection affecting both the short introns and 4D sites and to also correct for the approximation to the true SFS mentioned above. This allows the short intron SFS to not only act as neutral reference for the amplitude of the 4D SFS, but also its shape. With the correction, each frequency class now has a modifier, $\alpha_x$, which adjusts the probability of seeing a site with a SNP at frequency of *x* in the sample. As the $\alpha$'s are shared between the short intron and 4D SFS, they control for confounding factors affecting both spectra. This frequency-correction modifies equations (5) and (6) like so:

(7) $$\lambda_{full}\left(D \mid \theta, \overline{\gamma}, \overline{f}, \overline{\alpha}\right) = \lambda_{4D}\left(D \mid \theta, \overline{\gamma}, \overline{f}, \overline{\alpha}\right) \times \lambda_{SI}\left(D \mid \theta, \overline{\alpha}\right)$$

(8) $$\lambda_{SI}\left(D \mid \theta, \overline{\alpha}\right) = \prod_{x=0}^{x=\frac{1}{2}} \left(\alpha_x p(x \mid \theta)\right)^{k_x} where\ \alpha_0 = 1$$

While this correction is robust for many confounding factors [40], it adds a free parameter for every frequency-class except the first one. The parameter for the zero-frequency class, $\alpha_0$, is set to 1 to anchor the maximum-likelihood estimation of the $\alpha$'s. With 65 frequency classes, this adds 64 free parameters to the basic model of 5 free parameters.

**Phylogenetic tree and conservation**

We used the determined 15 species Insect tree topology from the UCSC genome browser (http://hgdownload.cse.ucsc.edu/goldenPath/dm3/phastCons15way/) and paired it down to the 12 Drosophila species [122]. We then input that tree topology into PhyML v3.0 (http://www.atgc-montpellier.fr/phyml) [69] and allowed it to re-estimate the branch lengths on all 4D sites in conserved amino acids using the HKY85 model [123] without a discrete gamma model and without invariant sites. The nucleotide frequencies and transition-transversion rate ratio were inferred by maximum-likelihood. The resulting tree can be found in the supplement (see Supplement – S8).

GERPcol from GERP++ (http://mendel.stanford.edu/SidowLab/downloads/gerp/) [72] was run on the collection of all 4D sites from all 12 Drosophila species excluding *D. melanogaster* and *D. wilistoni*, estimating the Rscore (tree length - inferred # of substitutions) for each site independently. We input into GERP the tree and transitition-transversion ratio from the PhyML results. As these two programs use different parameterizations of the transition-transversion ratio, we translated one to the other (see Supplement – S8).

**GO category enrichment**

Our signal from polymorphism does not afford us a precise measurement of constraint on the 4D sites of a single gene (not enough information). Therefore, we use a surrogate to infer the amount of strong constraint at the 4D sites of individual genes. Looking only at sites without SNPs, we use the percentage of 4D sites in conserved amino acids that are unpreferred and themselves conserved from *D. sechellia to D. grimshawi* (i.e. in the 0-substitution class) as our measure of how extensive the strong constraint has been on the 4D sites of the gene in question. As unpreferred 4D sites in the 0-substitution class have the highest fraction of sites under strong constraint (53%), the reasoning is that the more such sites exist in a gene, the more likely there has been extensive constraint acting on all 4D sites. Since not all genes have enough conserved amino acids to allow a reasonable calculation of the above surrogate, we used only those genes where at least 20% of the four-fold amino acids were conserved along the tree, leaving 4,877 genes in the analysis. We ranked genes by this surrogate and took the top 812 genes (~ top sixth of genes). We then used the functional annotation clustering tool from DAVID 6.7 (http://david.abcc.ncifcrf.gov/home.jsp) set on high stringency to look for enrichment of GO category terms in this gene set [85, 86].

**Acknowledgements**

The authors would like to thank David Enard, Anna-Sophie Fiston-Lavier, Penka Markova-Raina, Sandeep Venkataram, and all members of the Petrov lab for helpful feedback and support of this project.

**Funding**

The work was supported by the NIH grants RO1GM100366 and RO1GM097415 to DAP. RH is supported by an ERC FP7 CIG grant, by a Yigal Allon Fellowship awarded by the Israeli Council for Higher Education, and by the Robert J. Shillman Career Advancement Chair. PWM is supported by NIH grants RO1GM100366 and RO1GM097415. DSL is supported by the Stanford Genome Training Program (SGTP; NIH/NHGRI) and by NIH grants RO1GM100366 and RO1GM097415.

**Competing Interests**

The authors have no competing interests to report.

**Author Contributions**

Conceived and Designed Methodologies: DSL, PWM, RH, DAP. Wrote and Executed Programs: DSL. Analyzed results: DSL, PWM, RH, DAP. Wrote Paper: DSL, PWM, RH, DAP.

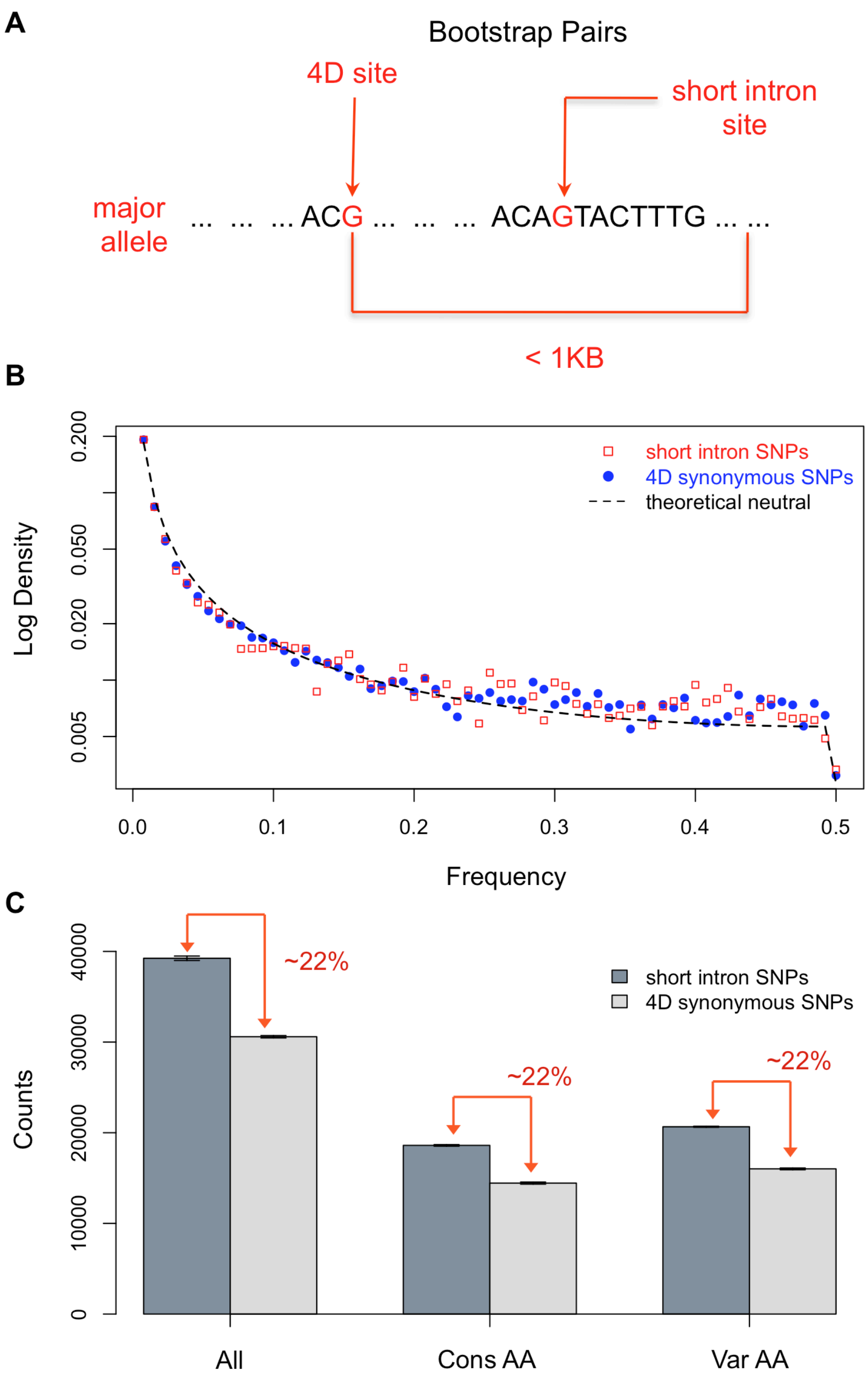

**Figure 1**

The signal of strong selection acting on 4D sites. (A) Overview of the bootstrap method. We sample 4D sites and their nearby (< 1KB apart) short intron pairs with replacement in order to control for linked selection and variation in GC content and mutation/recombination rates between the neutral reference (short introns) and the test set (4D sites). The short intron, 4D pair must have the same nucleotide as their major allele. (B) The folded Site Frequency Spectra (SFS) of observed SNPs from short introns, 4D sites, and the theoretical neutral distribution in a population with constant size. The SNPs were resampled to 130 strains and folded using the minor allele frequency. (C) The ratio of the amount of polymorphism in short introns versus 4D sites in all, conserved, and variable amino acids with standard error bars. Conserved amino acids are those present and identical in the 12 sequenced Drosophila genomes. Variable amino acids are defined as being not conserved according to the above definition. Ten bootstraps were done for each category (all, conserved, and variable) of 4D site. Lifting the restriction on distance and only controlling for GC content in the bootstrap produces identical results as above (not shown). To be conservative, we continued to use the distance restriction in the bootstrap. Note, had we simply taken the density of polymorphism as is without correction of GC content, we would've only seen a 7% drop in the density of polymorphism from short introns to 4D sites (5.58% vs 6.0% segregating in 4D vs short intron sites).

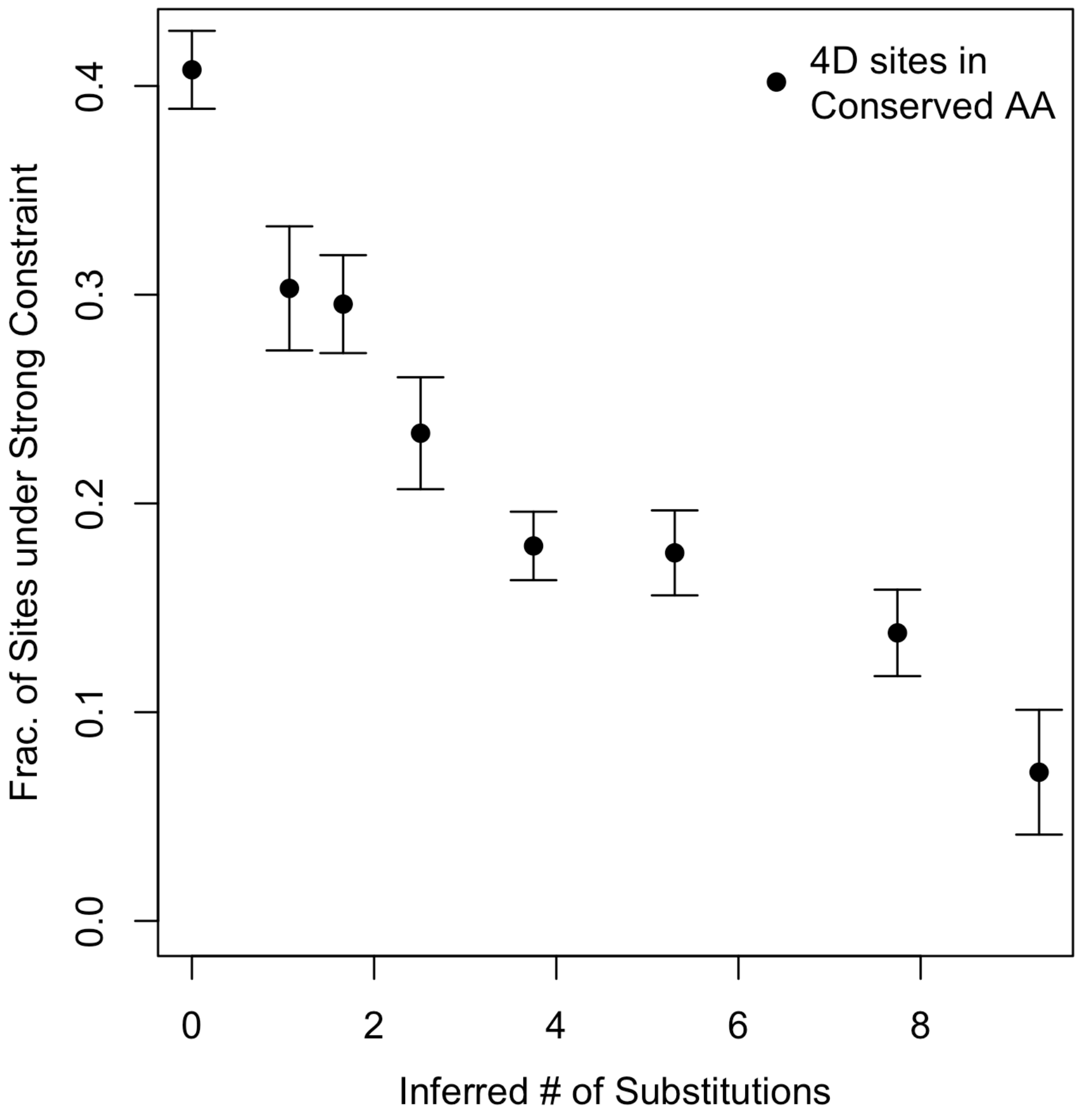

**Figure 2**

Conservation versus constraint at 4D sites in conserved amino acids. For each 4D site in a conserved amino acid, we use GERP to infer the number of substitutions that have occurred at that site across the Drosophila tree (removing *D. melanogaster* and *D. wilistoni* from the analysis). We define eight rate classes defined by the number of inferred substitutions across the tree - a proxy for the rate of evolution at the site - and bin the 4D sites accordingly. The class of the slowest evolving sites consists of those codons completely conserved across the ten Drosophila species (0 inferred substitutions along the tree at the 4D site). The fastest evolving class meanwhile has sites with greater than or equal to 9.3 substitutions per site. The remaining substitution classes are spread at intermediate values with a view to roughly equilibrate the number of sites in each class. The substitution bins (b) are as follows: ($b_1 = 0$, $0 < b_2 \leq 1.4$, $1.4 < b_3 \leq 1.92$, $1.92 < b_4 \leq 3.10$, $3.10 < b_5 \leq 4.40$, $4.40 < b_6 \leq 6.20$, $6.20 < b_7 < 9.30$, $b_8 \geq 9.30$). 10 bootstraps were done for the 4D sites within each bin and their short introns partners. Error bars represent the s.e. of the estimates.

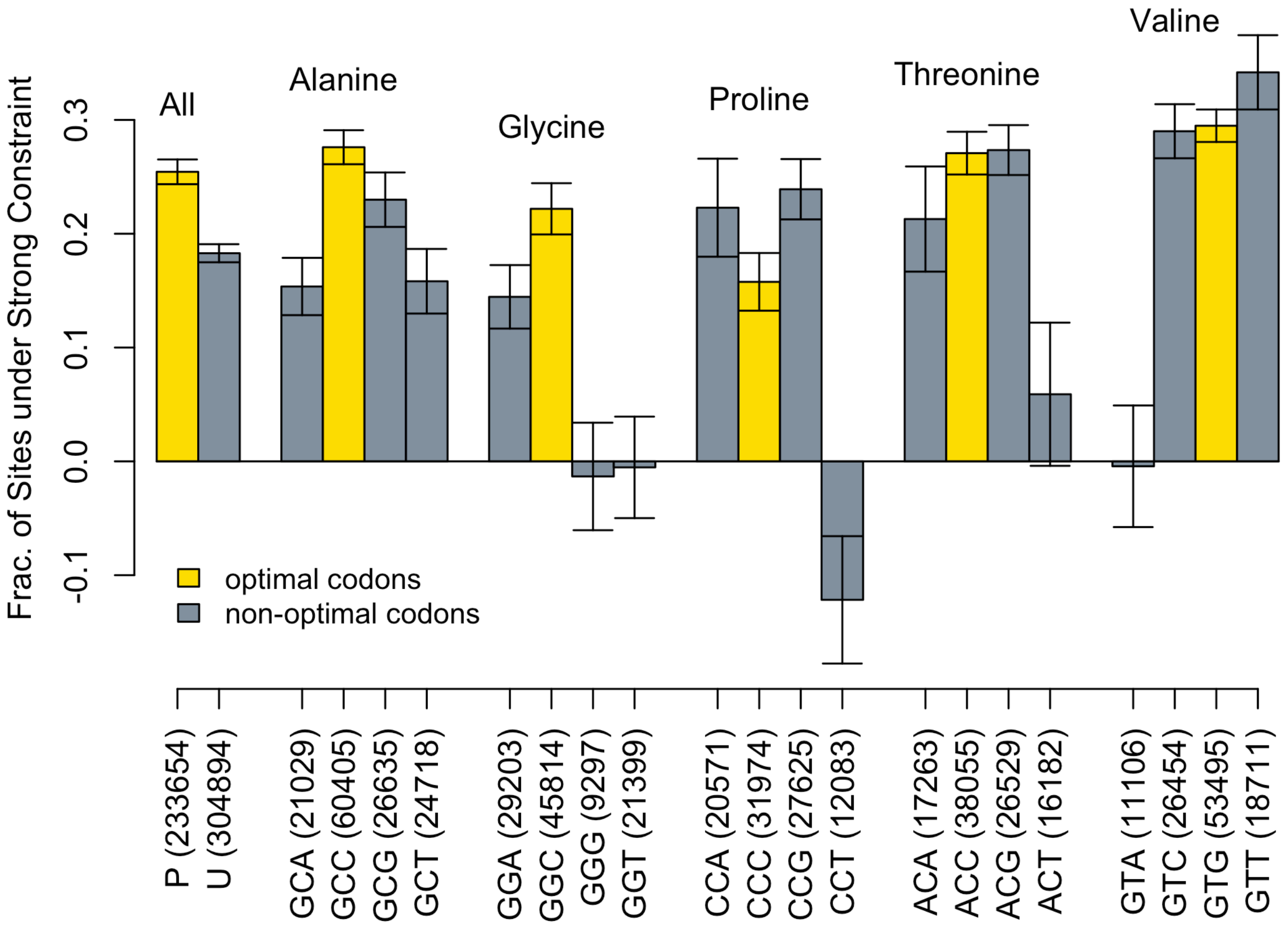

**Figure 3**

Constraint across codons. For each amino acid, we list the codons and, in parentheses, the number of 4D sites from each codon used in the bootstrap analysis – representing, in relative terms, the abundance of each codon in the genome. P-codons are all 4D sites from optimal codons grouped together, while U-codons are all 4D sites from non-optimal codons. 4D sites were binned into codons either by their ancestral allele as determined by parsimony to *D. sechellia* or by major allele if there is a substitution at that site between *D. sechellia* and *D. melanogaster*. Gold bars are the optimal codons for each amino acid, while dark grey bars are the non-optimal codons. 10 bootstraps determine the fraction of sites under constraint for each codon-type. Error bars represent the s.e. of the estimates. A negative value indicates an excess of polymorphism at 4D sites compared to short introns and is likely due to mispolarization assigning SNPs to the wrong codon.

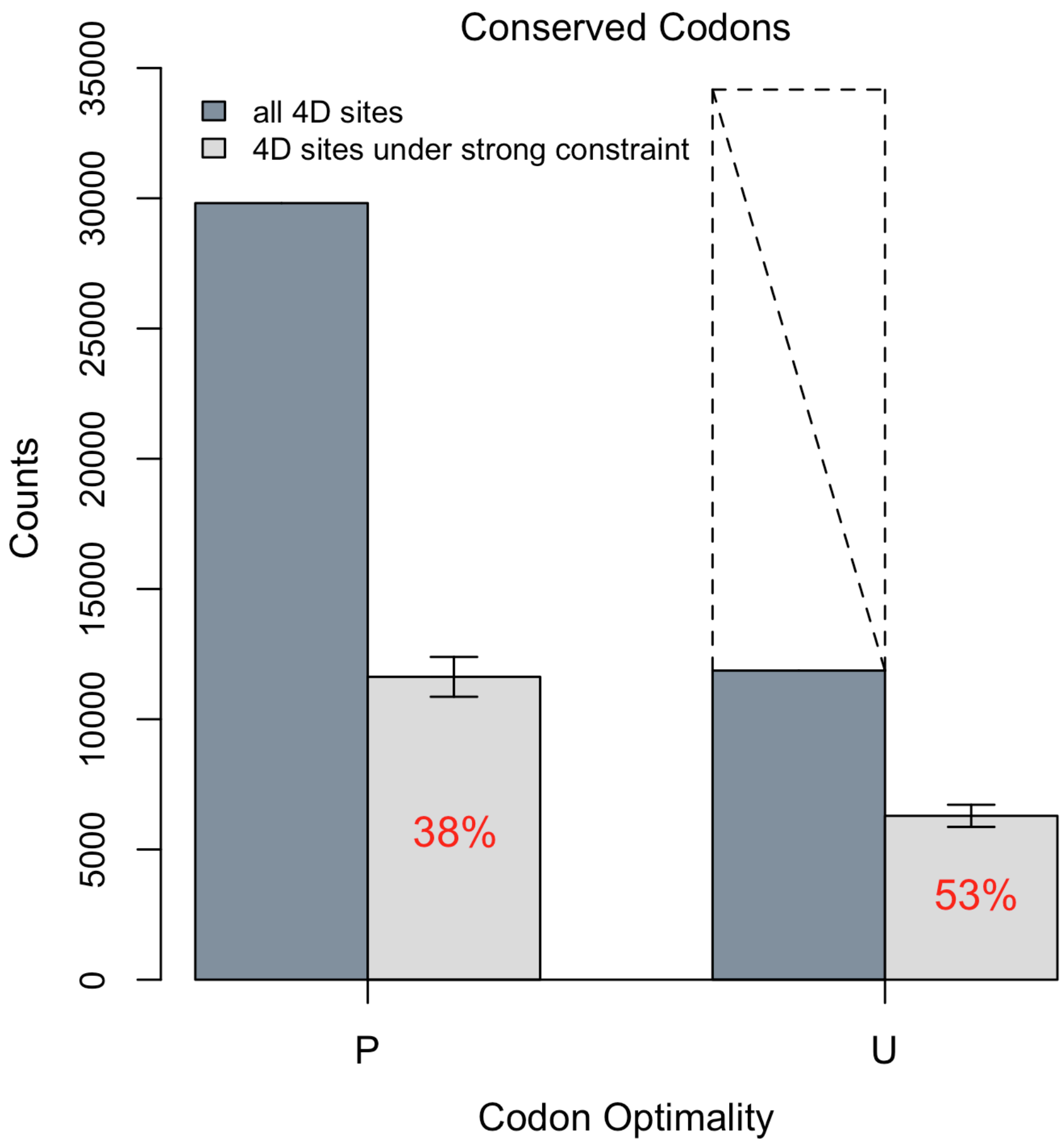

**Figure 4**

Codon optimality versus constraint in codons conserved from *D. sechellia-D. grimshawi* (excluding *D. willistoni*). The conserved codons were separated into those that were ancestrally preferred (P) and those that were ancestrally unpreferred (U) using polarization with the *D. sechellia-D. grimshawi* (excluding *D. willistoni*) outgroup. 10 bootstraps were done within each class. Error bars represent the s.e. of the estimates. The dark bars represent the counts of all sites that fall into each class while the light bars represent the number of sites estimated to be under strong constraint via the bootstrap procedure. The dashed line indicates what the count of total unpreferred conserved codons would have been had unpreferred 4D sites been conserved to the same extent as preferred 4D sites in otherwise conserved amino acids, i.e. the dashed line represents the proportion of U:P in all conserved amino acids. More than half (53%) of those unpreferred codons that are conserved across the ten Drosophila species are under strong purifying selection in *D. melanogaster*; 38% of preferred conserved codons are under strong selection.

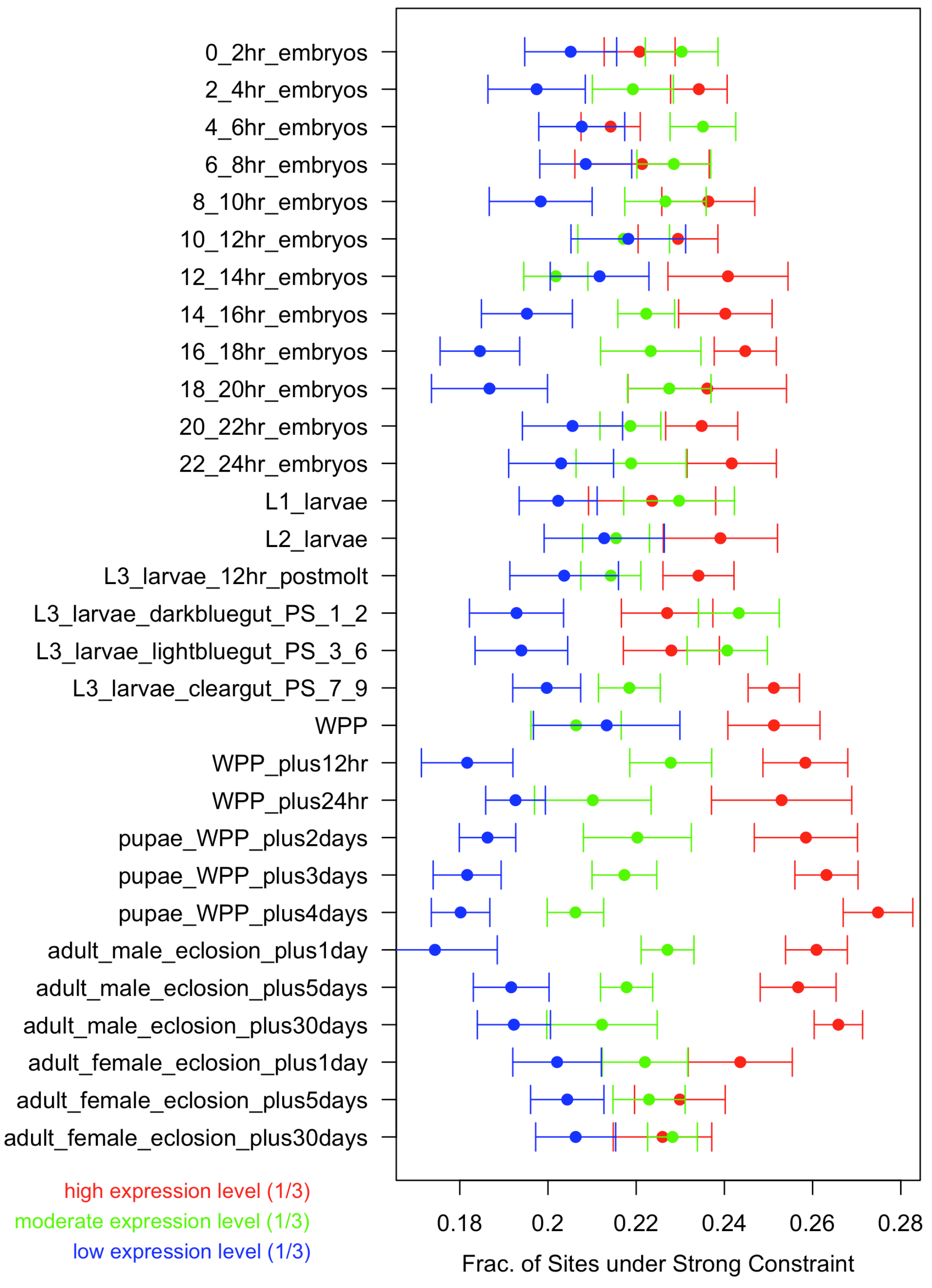

**Figure 5**

Strong constraint versus gene expression across development. Within each developmental time point, genes were ranked by their level of expression and then grouped into high, moderate, and low expression levels - each group comprising of one-third of all genes. Within each gene set within each time point, the fraction of 4D synonymous sites under strong constraint was calculated using the bootstrap. 10 bootstraps were done within each such class. Error bars represent the s.e. of the estimates.

**Table 1**

| **Selection Category**[a] | **Fraction of Sites**[b] | **Strength**[c] |
|---|---|---|
| Neutral | 77.4% (+/- 0.6%) | 0 |
| Weak Constraint | 0 | N/A |
| Strong Constraint | 22.6% (+/- 0.6%) | -283 (+/- 28.3) |

[a]selection categories are defined as follows => Neutral: $4N_es = 0$, Weak Constraint: $|4N_es| < 5$, and Strong Constraint: $|4N_es| > 100$ (defining Strong Constraint: $|4N_es| > 5$ gives exactly the same MLE for the fraction/strength of the strong category); [b]mean of the MLEs for the fraction of sites in each category over the ten bootstrap runs (+/- s.e.); [c]mean of the MLEs for the strength of strong selection over the ten bootstrap runs (+/- s.e.); $4N_e\mu$ ($\theta$) = 0.0132

**Table 2**

| FOP[a] | Fraction of Sites[c] | ENC[b] | Fraction of Sites[c] |
|---|---|---|---|
| high FOP | 18.7% (+/- 1.7%) | low ENC | 21.8% (+/- 1.5%) |
| medium FOP | 23.1% (+/- 1.0%) | medium ENC | 21.8% (+/- 0.8%) |
| low FOP | 23.2% (+/- 0.9%) | high ENC | 23.0% (+/- 1.0%) |

[a]genes are ranked in descending order by their FOP with the the top, middle, and bottom third forming the high, medium, and low FOP classes respectively; [b]genes are ranked in ascending order by their ENC with the the top, middle, and bottom third forming the low, medium, and high ENC classes respectively; [c]mean fraction of 4D sites under strong constraint in each category over 10 bootstrap runs (+/- s.e.)

**Table 3**

| Category | Fraction of Sites[a] | Category | Fraction of Sites[a] |
|---|---|---|---|
| 5’ 75bp of CDS[b] | 30.7% (+/- 3.0%) | 3’ 75bp of CDS[c] | 31.5% (+/- 2.5%) |
| Bulk Nucleosomes[d] | 24.2% (+/- 0.7%) | splice junctions[e] | 26.0% (+/- 1.0%) |
| multi-exon genes[f] | 22.0% (+/- 0.6%) | single-exon genes[g] | 21.8% (+/- 4.4%) |
| long genes[h] | 25.8% (+/- 0.9%) | long CDSs[i] | 24.1% (+/- 0.8%) |
| medium genes[h] | 19.3% (+/- 0.6%) | medium CDSs[i] | 19.2% (+/- 1.0%) |
| short genes[h] | 17.3% (+/- 1.3%) | short CDSs[i] | 20.3% (+/- 1.8%) |
| autosomal genes[j] | 22.6% (+/- 0.5%) | X-linked genes[k] | 19.2% (+/- 1.3%) |

[a]mean fraction of 4D sites under strong constraint in each category over 10 bootstrap runs (+/- s.e.); [b]4D sites within 75bp of the translation start site (longest transcript); [c]4D sites within 75bp of stop codon (longest transcript); [d]4D sites in bulk nucleosome footprints; [e]4D sites within 48bp of a splice junction; [f]4D sites from multi-exon genes; [g]4D sites from single-exon genes; [h]genes are ranked in descending order by their gene length (UTR + all exons + all introns) with the the top, middle, and bottom third forming the long, medium, and short gene classes respectively; [i]genes are ranked in descending order by their CDS length (longest transcript) with the the top, middle, and bottom third forming the long, medium, and short CDS classes respectively; [j]4D sites from Autosomal genes; [k]4D sites from X-linked genes.

**Table 4**

| Cluster #[a] | Overall Functional Annotation[b] | Enrichment[c] |
|---|---|---|
| 1 | transcriptional regulation | 9.69 |
| 2 | imaginal disc development | 9.28 |
| 3 | homeobox protein domain | 7.57 |
| 4 | eye morphogenesis | 7.49 |
| 6 | epithelium development | 6.07 |
| 8 | immunoglobulin domain | 5.93 |
| 9 | ribosomal proteins | 5.36 |
| 10 | cell signaling | 4.59 |
| 12 | gamete generation | 4.33 |
| 13 | neuron development | 3.51 |

[a]Functional annotation clusters ranked by significance by DAVID 6.7 [85, 86]. These clusters are groups of similar or related biological annotation terms, with similarity determined by a simple stringency setting - in the above, a high stringency setting was used. The significance of the overall cluster reflects the combined enrichment in the test gene set of the individual annotation terms within a cluster (see c). Clusters 5, 7, and 11 are not reported here as their biological terms were similar to clusters 4, 1, and 4 & 13 respectively, so provided no new information. The full information for the top 13 clusters is reported in the supplement (see Supplement – S6); [b]Summary description of the type of annotation terms within each cluster. The specific annotation terms for each cluster are in the supplement; [c]The enrichment score of the overall cluster as calculated by DAVID in the test gene set with respect to the background gene set. According to the description of enrichment scores by DAVID, each individual annotation term within a cluster has a p-value, or significance, for the enrichment of that term in the test gene set. The enrichment score of the overall cluster is then the geometric mean of these p-values. Thus the higher the enrichment score, the lower the p-values are for all terms in the overall annotation cluster and the more significantly enriched the overall cluster is in the test gene set. The p-values for the enrichment of the annotation terms in each cluster are reported in the supplement (see Supplement – S7).

**S1 – Demographic correction of SFS**

We used a frequency-dependent variable $\alpha$ to correct for demography, weak selection affecting both test and reference sites, and our approximation to the true SFS (see Materials and Methods). The frequency correction adds 64 additional free variables over which the likelihood needs to be maximized (see Materials and Methods). Using fminsearch in Matlab as before to perform the maximum-likelihood (see Materials and Methods), we ran several different variable initializations of f (fraction of sites in each category), $\gamma$ (strength of selection, $4N_e s$, in each category), and $\theta$ (effective mutation rate, $4N_e\mu$) for each of the 10 bootstraps and then chose the local maxima with the lowest log likelihood out of the initializations. These reported values thus cannot be guaranteed to be the global maxima for each bootstrap.

f = ($f_{neutral}$, $f_{weak}$ , $f_{strong}$); gamma = ($\gamma_{weak}$, $\gamma_{strong}$); theta
0) f = (0.7832, 0.0086, 0.2082); $\gamma$ = (-0.7391, -531.2); $\theta$ = 0.0137
1) f = (0.7682, 0.0046, 0.2272); $\gamma$ = (-3.8237, -497.0); $\theta$ = 0.0140
2) f = (0.7710, 0.0094, 0.2196); $\gamma$ = (-3.6057, -332.4); $\theta$ = 0.0135
3) f = (0.7642, 0.0107, 0.2251); $\gamma$ = (-4.6512, -324.5); $\theta$ = 0.0137
4) f = (0.7697, 0.0010, 0.2293); $\gamma$ = (-0.0018, -295.1); $\theta$ = 0.0137
5) f = (0.7667, 0.0089, 0.2244); $\gamma$ = (-0.1215, -336.1); $\theta$ = 0.0137
6) f = (0.7784, 0.0106, 0.2110); $\gamma$ = (-4.6003, -351.5); $\theta$ = 0.0138
7) f = (0.7614, 0.0090, 0.2296); $\gamma$ = (-0.0003, -310.8); $\theta$ = 0.0141
8) f = (0.7677, 0.0128, 0.2195); $\gamma$ = (-4.0697, -507.4); $\theta$ = 0.0138
9) f = (0.7475, 0.0231, 0.2294); $\gamma$ = (-0.0002, -215.2); $\theta$ = 0.0136

avg % of sites under strong selection: 22.23% +/- 0.767 (+/- s.e.)
avg strength of strong selection: -370.12 +/- 105 (+/- s.e.)

As compared to the estimates without demographic correction, the percent of sites under strong selection is roughly the same, but the strength of that selection is somewhat higher. The variance of that latter estimation is unsurprisingly larger given the increased number of variables. The estimated intensity of the strong constraint is however still greater than three times the standard error away from both -700, the calculable limit of our program, and weak selection (-5).

**S2 – Mutation versus Strong Constraint**

*S2.A – Tri-nucleotide bootstrap results*

To test if the lack of polymorphism at 4D sites relative to short introns could be explained by a lower context-dependent mutation rate at 4D sites, we ran bootstraps matching the 4D sites and their flanking nucleotides with nearby (again, < 1KB) triplets of short intron sites. We split the results up by whether the 4D sites are in conserved versus variable amino acids (conservation is across the 12 Drosophila species tree). In the 4D sites of otherwise conserved amino acids, 18.5% (+/- 1.5%) (+/- s.e.) of sites are missing polymorphism. For 4D sites in variable amino acids, the percentage of missing polymorphism drops to 16.6% (+/- 1.1%). So matching triplets to control for dinucleotide content would appear to explain some fraction of the overall signal of missing polymorphism. However, closer analysis reveals this to be an artifact of the type of 4D site capable of matching nearby short introns when controlling for neighbors. The GC content of all 4D sites falling in conserved and variable amino acids is 67.1% and 61.9% respectively, while the GC content of those 4D sites capable of matching neighbors with short introns is ~61.3% and ~55.8% respectively. The GC content of short introns is far lower than that of 4D sites and thus 4D sites with G or C as their major allele have more trouble finding nearby short intron G/C sites with the same flanking nucleotides. Thus the makeup of the 4D sites capable of being used in the in the tri-nucleotide bootstrap is skewed compared to the full sample. Just taking those 4D sites from the tri-nucleotide bootstrap and matching them instead to short intron sites with the same major allele, but not the same flanking nucleotides reveals a nearly identical amount of missing polymorphism to the above: 19.5% (+/- 0.9%) and 16.6% (+/- 0.9%) for 4D sites in conserved and variable amino acids respectively. The matched short intron sites now have a different neighborhood than the 4D sites, but the resulting drop in polymorphism is the same. In the triplet bootstrap, the sample of 4D sites is biased by the neighborhood control rather than the neighborhood controlling for a context-dependent mutation rate. As such, dinucleotide biases and consequent context-dependent mutational effects do not explain any appreciable amount of the drop in polymorphism in 4D sites relative to short introns.

*S2.B – Power to detect low mutation versus strong purifying selection on 4D sites*

We simulated three sample spectra matching the total number of sites and the depth of population sample to that observed in the data: a neutral reference SFS, a low-mutation SFS, and a strong constraint SFS (see Materials and Methods). We used the first as a reference against which to measure the apparent amount of selection acting on the latter two. For the neutral, but 22%-lower-mutation-rate SFS, we estimated the strength of effective selection to be $4N_es \sim -700$ on 22% of sites, the calculable limit of our program for Drosophila-like parameters and thus essentially infinitely strong purifying selection. In the third SFS, 22% of the sites evolved under a constraint of $4N_es = -283$ with the rest neutral. Re-estimating the intensity of the strong constraint category to be -283 provides a significantly better fit to the SFS than setting its value equal to -700 *a priori* (p-value: 0.0366, LRT ~ $X^2_1$). When the strong selection category is set *a priori* to be infinitely strong, the maximum-likelihood procedure compensates for the slight differences between the short intron and 4D site spectra with small amounts of weak selection. If we remove the weak selection category from the maximum-likelihood analysis, the above likelihood differential increases in favor of a finite strong selective force over mutational force explaining the SFS (p-value: 0.0137, LRT ~ $X^2_1$). So while weak selection can buoy the likelihood of a mutational force explaining the spectra, we have enough sites from across the genome and samples from within a population to distinguish between a SFS under a finite, strong selective force and a low-mutation SFS.

*S2.C – Slow- versus fast-evolving 4D site bootstrap results*

Using the same intervals as Figure 2, we can use fastest evolving 4D sites in otherwise conserved amino acids, class $b_8$ in Figure 2, as the neutral reference against which to measure the amount of missing polymorphism in 4D sites across the different substitution-rate classes. We matched slow-evolving and fast-evolving 4D sites as before by major allele, controlling for distance (< 1KB). When matching fast-evolving 4D sites to each other in the bootstrap, we prohibited a 4D site from matching with itself. As shown in Figure 2, the fast-evolving 4D sites are not a perfect neutral reference – a few are themselves under strong constraint. The percent of missing polymorphism in fast-evolving 4D sites used the neutral reference for each rate-class were measured relative to short introns. Overall it is 7.1% (+/- 3.0%).

| **Rate Class**[a] | **Fraction of Sites**[b] | **Fraction of Sites**[c] | **Fraction of Sites**[d] |
|---|---|---|---|
| $b_1 = 0$ | 27.1% (+/- 1.6%) | 13.7% (+/- 3.3%) | 40.1% (+/- 1.9%) |
| $0 < b_2 \le 1.4$ | 21.8% (+/- 2.3%) | 9.2% (+/- 3.4%) | 30.3% (+/- 3.0%) |
| $1.4 < b_3 \le 1.92$ | 18.1% (+/- 2.6%) | 9.0% (+/- 3.6%) | 29.5% (+/- 2.3%) |
| $1.92 < b_4 \le 3.10$ | 17.2% (+/- 2.6%) | 4.7% (+/- 3.1%) | 23.4% (+/- 2.7%) |
| $3.10 < b_5 \le 4.40$ | 10.7% (+/- 2.5%) | 10.3% (+/- 2.5%) | 18.0% (+/- 1.6%) |
| $4.40 < b_6 \le 6.20$ | 11.3% (+/- 2.4%) | 8.8% (+/- 2.5%) | 17.6% (+/- 2.0%) |
| $6.20 < b_7 < 9.30$ | 3.6% (+/- 2.4%) | 10.8% (+/- 2.4%) | 13.8% (+/- 2.1%) |
| $b_8 \ge 9.30$ | 0.7% (+/- 2.4%) | 6.5% (+/- 3.2%) | 7.1% (+/- 3.0%) |

[a]4D sites in otherwise conserved amino acids classified by substitution rate; [b]mean fraction of sites missing polymorphism over 10 bootstrap runs in 4D sites of each rate-class using fast-evolving 4D sites as reference (+/- s.e.); [c]mean fraction of sites missing polymorphism relative to nearby short intron sites in those fast-evolving 4D sites used as the neutral reference for each rate-class; [d]mean fraction of sites missing polmyorphism in each rate class (short introns as neutral reference, same as reported in Figure 2).

Summing the two columns of missing polymorphism together (b,c) almost perfectly recapitulates the fraction of sites under strong constraint as measured in Figure 2 (d).

Applying our SFS model (w/o frequency-dependent correction) to test all slow-evolving 4D sites (< 9.3) for selection, using fast-evolving 4D sites (>= 9.3) as neutral reference, yields the following result:

| **Selection Category**[a] | **Fraction of Sites**[b] | **Strength**[c] |
|---|---|---|
| Neutral | 84.2% (+/- 0.7%) | 0 |
| Weak Constraint | 0 | N/A |
| Strong Constraint | 15.8% (+/- 0.7%) | -307 (+/- 105) |

[a]selection categories are defined as follows => Neutral: $4N_es = 0$, Weak Constraint: $|4N_es| < 5$, and Strong Constraint: $|4N_es| > 100$ (defining Strong Constraint: $|4N_es| > 5$ gives exactly the

same MLE for the fraction/strength of the strong category); [b]mean of the MLEs for the fraction of slow-evolving 4D sites in each category over the ten bootstrap runs (+/- s.e.); [c]mean of the MLEs for the strength of strong selection over the ten bootstrap runs (+/- s.e.); $4N_e\mu$ ($\theta$) = 0.0124

The above shows that our results are not dependent on using short introns as a neutral reference – that they are not due to any artifact of short introns. Remaining selection on fast-evolving 4D sites results in a more conservative estimate of the fraction of sites under constraint when they are used as the "neutral" reference, but regardless there is less polymorphism at slower-evolving 4D sites than fast-evolving 4D sites and a signal of strong constraint in their relative site-frequency spectra. Further tests were done matching neighborhoods of slow and fast evolving 4D sites in the bootstrap and as in S2.A, no significant effects on the amount of missing polymorphism were found (not shown).

**S3 – Spatial distribution of strong constraint within coding sequences.**

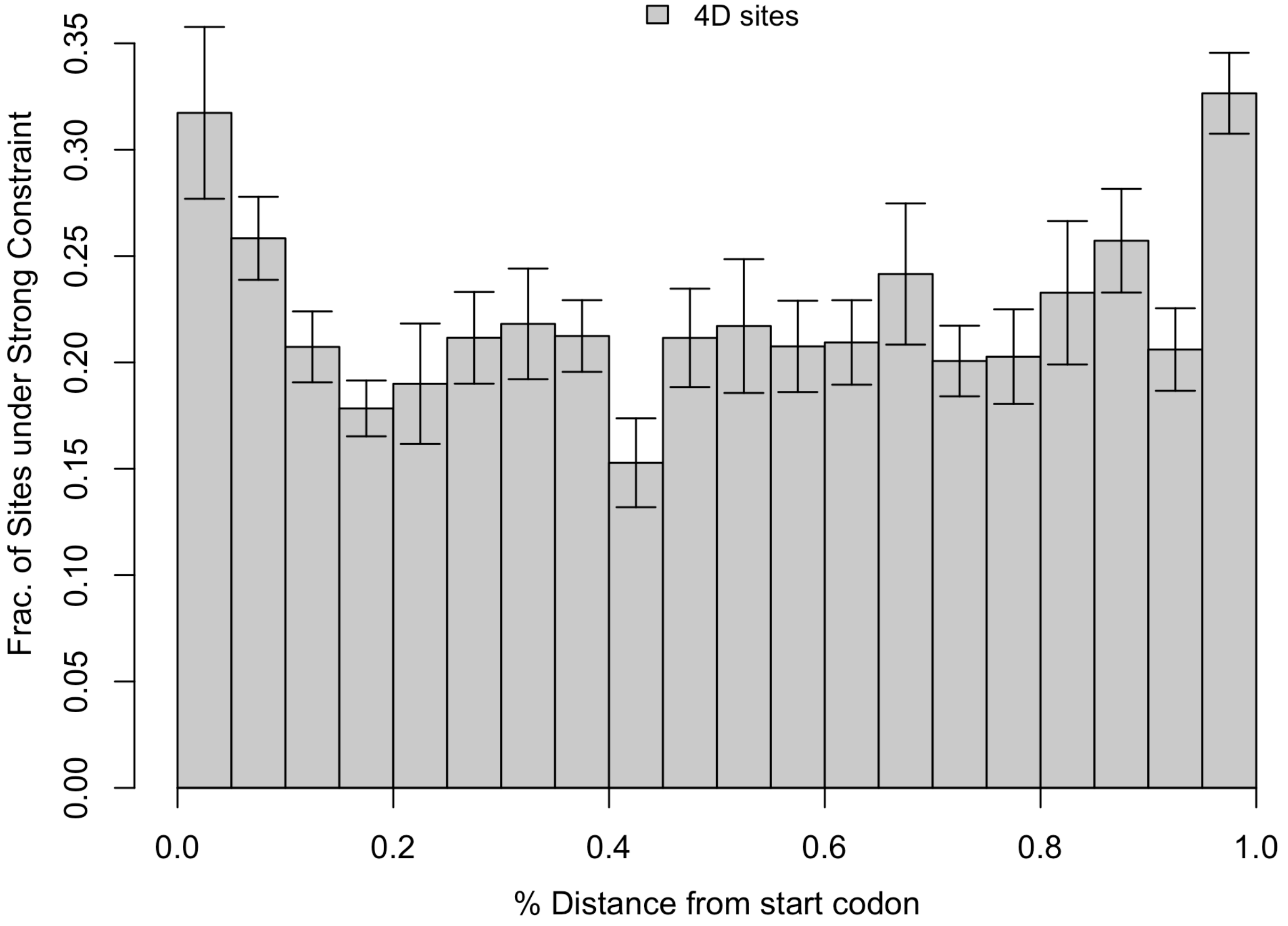

4D sites were binned by their distance to the translation start site in the longest transcript for each gene. Each bin represents 5% of transcript length to control for different transcript lengths. 10 bootstraps to determine the fraction of sites under constraint were done within each bin. Error bars represent the s.e. of the estimates.

**S4 – Weak Selection on Bulk Nucleosomes**

Bulk nucleosomes cover about 67% of our 4D sites. However, 55% of our short introns positions are also bound by nucleosomes. Comparing the level of polymorphism in nucleosomal to non-nucleosomal short intron positions, we see a 9.0% drop even before accounting for the GC differences between them. Further the minor allele frequencies of SNPs in nucleosomal short intron positions is lower than those of SNPs in non-nucleosomal short intron positions (Supplemental). As such, while 24% of polymorphism may be missing in 4D sites bound nucleosomes relative to all short intron positions, the fraction of sites under strong constraint is still truly ~22%. Our previous SFS analysis showed no sign of weak selection on 4D sites, but bulk nucleosomes cover short introns to almost the same extent as 4D sites. The signal from any selective or mutational force caused by nucleosomes is washed out when comparing 4D sites with short introns. Thus the binding of bulk nucleosomes does not explain the signal of strong constraint in 4D sites when compared to short introns. We did not see a correlation of strong constraint with active, H2A.Z nucleosomal or PolII binding (not shown).

**S5 – Expression level over development with 4 categories**

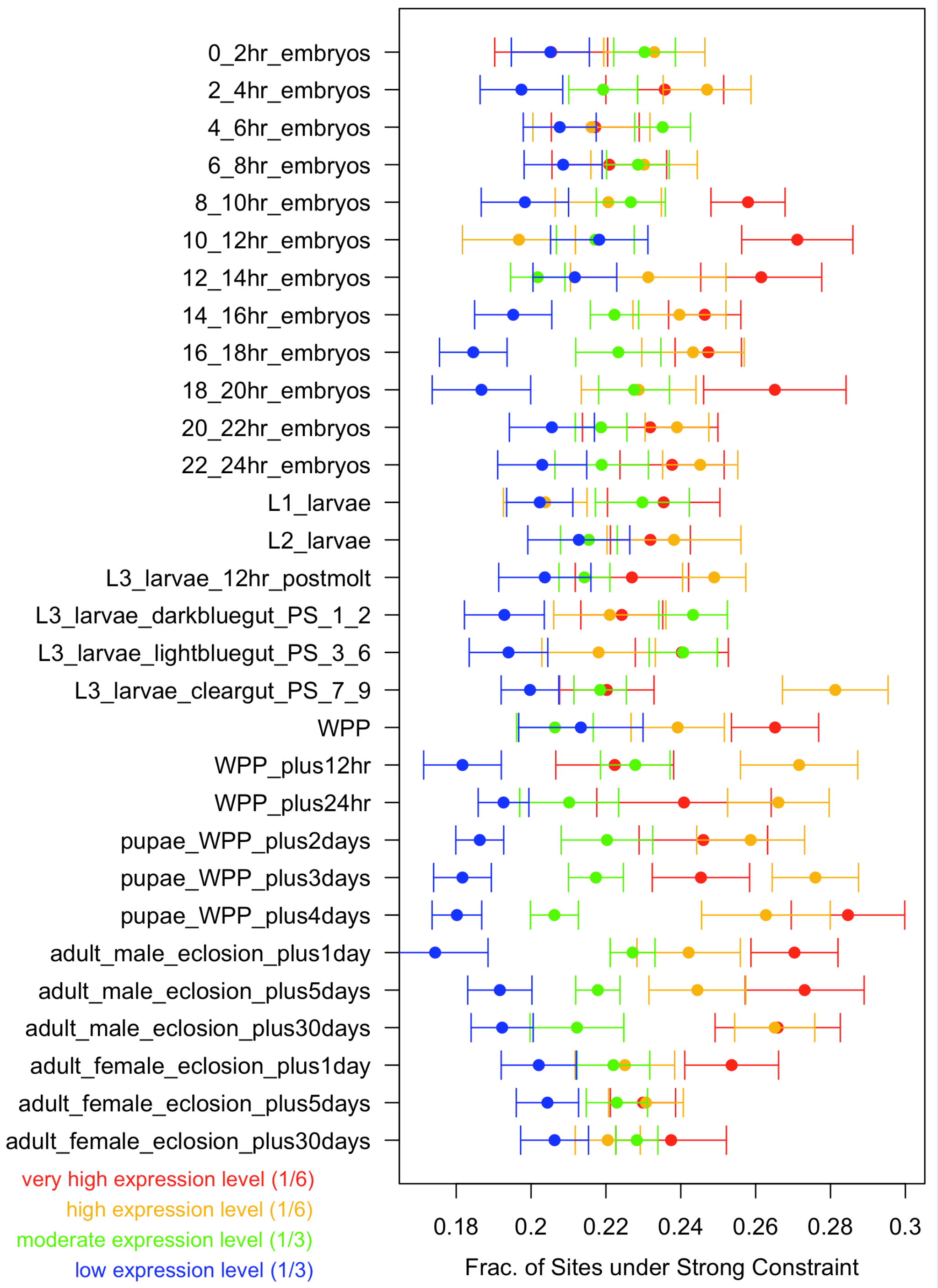

Strong constraint versus gene expression across development. Genes are grouped and analyzed as in Figure 5. We split the Figure 5 “high expression level” gene set in half into “very high expression level” and “high expression level” groups, thus each containing one-sixth of all genes.

## S6 – Gene ontology clusters

Below is the full information for the top 13 GO clusters as reported by DAVID 6.7 from the 812 genes most enriched for strong constraint at 4D sites. (citation)

**Annotation Cluster 1** **Enrichment Score: 9.691589451545648**

| Category | Term | Count | % | PValue | List Total | Pop Hits | Pop Total | Fold Enrichment | FDR |
|---|---|---|---|---|---|---|---|---|---|
| GOTERM_MF_FAT | GO:0030528~transcription regulator activity | 103 | 12.68472906 | 2.52E-11 | 536 | 323 | 3078 | 1.831211589 | 3.74E-08 |
| GOTERM_BP_FAT | GO:0006355~regulation of transcription, DNA-dependent | 90 | 11.08374384 | 3.88E-11 | 511 | 260 | 2817 | 1.908249285 | 6.67E-08 |
| GOTERM_BP_FAT | GO:0051252~regulation of RNA metabolic process | 95 | 11.69950739 | 6.10E-10 | 511 | 293 | 2817 | 1.787400733 | 1.05E-06 |
| GOTERM_BP_FAT | GO:0045449~regulation of transcription | 105 | 12.93103448 | 2.87E-09 | 511 | 344 | 2817 | 1.682661676 | 4.95E-06 |

**Annotation Cluster 2** **Enrichment Score: 9.281593760361073**

| Category | Term | Count | % | PValue | List Total | Pop Hits | Pop Total | Fold Enrichment | FDR |
|---|---|---|---|---|---|---|---|---|---|
| GOTERM_BP_FAT | GO:0007444~imaginal disc development | 69 | 8.497536946 | 7.45E-14 | 511 | 159 | 2817 | 2.392312521 | 1.28E-10 |
| GOTERM_BP_FAT | GO:0002165~instar larval or pupal development | 70 | 8.620689655 | 2.65E-11 | 511 | 180 | 2817 | 2.143835616 | 4.56E-08 |
| GOTERM_BP_FAT | GO:0048563~post-embryonic organ morphogenesis | 52 | 6.403940887 | 5.66E-11 | 511 | 117 | 2817 | 2.450097847 | 9.75E-08 |
| GOTERM_BP_FAT | GO:0007560~imaginal disc morphogenesis | 52 | 6.403940887 | 5.66E-11 | 511 | 117 | 2817 | 2.450097847 | 9.75E-08 |
| GOTERM_BP_FAT | GO:0009791~post-embryonic development | 70 | 8.620689655 | 6.47E-11 | 511 | 183 | 2817 | 2.10869077 | 1.11E-07 |
| GOTERM_BP_FAT | GO:0048569~post-embryonic organ development | 52 | 6.403940887 | 1.76E-10 | 511 | 120 | 2817 | 2.388845401 | 3.04E-07 |
| GOTERM_BP_FAT | GO:0007552~metamorphosis | 61 | 7.512315271 | 3.88E-10 | 511 | 155 | 2817 | 2.169522126 | 6.67E-07 |
| GOTERM_BP_FAT | GO:0048737~imaginal disc-derived appendage development | 44 | 5.418719212 | 7.13E-10 | 511 | 96 | 2817 | 2.526663405 | 1.23E-06 |
| GOTERM_BP_FAT | GO:0035114~imaginal disc-derived appendage morphogenesis | 44 | 5.418719212 | 7.13E-10 | 511 | 96 | 2817 | 2.526663405 | 1.23E-06 |
| GOTERM_BP_FAT | GO:0009886~post-embryonic morphogenesis | 60 | 7.389162562 | 8.93E-10 | 511 | 154 | 2817 | 2.147813048 | 1.54E-06 |
| GOTERM_BP_FAT | GO:0035107~appendage morphogenesis | 44 | 5.418719212 | 1.07E-09 | 511 | 97 | 2817 | 2.500615329 | 1.84E-06 |
| GOTERM_BP_FAT | GO:0048736~appendage development | 44 | 5.418719212 | 1.07E-09 | 511 | 97 | 2817 | 2.500615329 | 1.84E-06 |
| GOTERM_BP_FAT | GO:0048707~instar larval | 58 | 7.142857143 | 1.90E-09 | 511 | 149 | 2817 | 2.145891068 | 3.27E-06 |
| | or pupal morphogenesis | | | | | | | | |
| GOTERM_BP_FAT | GO:0035120~post-embryonic appendage morphogenesis | 40 | 4.926108374 | 1.00E-08 | 511 | 89 | 2817 | 2.477627037 | 1.72E-05 |
| GOTERM_BP_FAT | GO:0035220~wing disc development | 44 | 5.418719212 | 1.45E-08 | 511 | 104 | 2817 | 2.332304682 | 2.50E-05 |
| GOTERM_BP_FAT | GO:0007476~imaginal disc-derived wing morphogenesis | 37 | 4.556650246 | 5.29E-08 | 511 | 83 | 2817 | 2.45747766 | 9.11E-05 |
| GOTERM_BP_FAT | GO:0007472~wing disc morphogenesis | 37 | 4.556650246 | 7.72E-08 | 511 | 84 | 2817 | 2.428221974 | 1.33E-04 |

**Annotation Cluster 3** **Enrichment Score: 7.573942674767135**

| Category | Term | Count | % | PValue | List Total | Pop Hits | Pop Total | Fold Enrichment | FDR |
|---|---|---|---|---|---|---|---|---|---|
| INTERPRO | IPR017970:Homeobox, conserved site | 26 | 3.201970443 | 6.85E-09 | 691 | 47 | 4190 | 3.354373865 | 1.08E-05 |
| INTERPRO | IPR001356:Homeobox | 26 | 3.201970443 | 6.85E-09 | 691 | 47 | 4190 | 3.354373865 | 1.08E-05 |
| SP_PIR_KEYWORDS | Homeobox | 26 | 3.201970443 | 1.23E-08 | 786 | 48 | 4770 | 3.28721374 | 1.63E-05 |
| INTERPRO | IPR012287:Homeodomain-related | 25 | 3.078817734 | 6.70E-08 | 691 | 48 | 4190 | 3.158164496 | 1.05E-04 |
| SMART | SM00389:HOX | 26 | 3.201970443 | 3.48E-07 | 377 | 47 | 1869 | 2.742479824 | 4.34E-04 |

**Annotation Cluster 4** **Enrichment Score: 7.4921762056957855**

| Category | Term | Count | % | PValue | List Total | Pop Hits | Pop Total | Fold Enrichment | FDR |
|---|---|---|---|---|---|---|---|---|---|
| GOTERM_BP_FAT | GO:0048592~eye morphogenesis | 41 | 5.049261084 | 5.74E-09 | 511 | 91 | 2817 | 2.483753038 | 9.87E-06 |
| GOTERM_BP_FAT | GO:0007423~sensory organ development | 58 | 7.142857143 | 8.07E-09 | 511 | 154 | 2817 | 2.07621928 | 1.39E-05 |
| GOTERM_BP_FAT | GO:0001745~compound eye morphogenesis | 37 | 4.556650246 | 5.29E-08 | 511 | 83 | 2817 | 2.45747766 | 9.11E-05 |
| GOTERM_BP_FAT | GO:0001654~eye development | 47 | 5.78817734 | 7.36E-08 | 511 | 120 | 2817 | 2.159148728 | 1.27E-04 |
| GOTERM_BP_FAT | GO:0048749~compound eye development | 44 | 5.418719212 | 1.92E-07 | 511 | 112 | 2817 | 2.16571149 | 3.30E-04 |

**Annotation Cluster 5** **Enrichment Score: 6.17076557737113**

| Category | Term | Count | % | PValue | List Total | Pop Hits | Pop Total | Fold Enrichment | FDR |
|---|---|---|---|---|---|---|---|---|---|
| GOTERM_BP_FAT | GO:0001751~compound eye photoreceptor cell differentiation | 27 | 3.325123153 | 5.25E-07 | 511 | 55 | 2817 | 2.70624444 | 9.05E-04 |
| GOTERM_BP_FAT | GO:0046530~photoreceptor cell differentiation | 28 | 3.448275862 | 7.16E-07 | 511 | 59 | 2817 | 2.616206176 | 0.001232255 |
| GOTERM_BP_FAT | GO:0001754~eye photoreceptor cell differentiation | 27 | 3.325123153 | 8.17E-07 | 511 | 56 | 2817 | 2.657918647 | 0.001406528 |

**Annotation Cluster 6** **Enrichment Score: 6.070992847167416**

| Category | Term | Count | % | PValue | List Total | Pop Hits | Pop Total | Fold Enrichment | FDR |
|---|---|---|---|---|---|---|---|---|---|
| GOTERM_BP_FAT | GO:0048729~tissue morphogenesis | 40 | 4.926108374 | 1.75E-07 | 511 | 97 | 2817 | 2.273286662 | 3.02E-04 |
| GOTERM_BP_FAT | GO:0060429~epithelium development | 37 | 4.556650246 | 8.56E-07 | 511 | 91 | 2817 | 2.241435668 | 0.001473044 |
| GOTERM_BP_FAT | GO:0002009~morphogenesis of an epithelium | 34 | 4.187192118 | 4.08E-06 | 511 | 85 | 2817 | 2.205088063 | 0.007028429 |

**Annotation Cluster 7** **Enrichment Score: 5.950492315603639**

| Category | Term | Count | % | PValue | List Total | Pop Hits | Pop Total | Fold Enrichment | FDR |
|---|---|---|---|---|---|---|---|---|---|
| SP_PIR_KEYWORDS | transcription regulation | 63 | 7.75862069 | 7.48E-08 | 786 | 197 | 4770 | 1.940752509 | 9.92E-05 |
| SP_PIR_KEYWORDS | Transcription | 63 | 7.75862069 | 2.07E-07 | 786 | 202 | 4770 | 1.892714081 | 2.75E-04 |
| GOTERM_BP_FAT | GO:0006350~transcription | 66 | 8.128078818 | 9.07E-05 | 511 | 234 | 2817 | 1.554869788 | 0.156037459 |

**Annotation Cluster 8** **Enrichment Score: 5.930318493377584**

| Category | Term | Count | % | PValue | List Total | Pop Hits | Pop Total | Fold Enrichment | FDR |
|---|---|---|---|---|---|---|---|---|---|
| INTERPRO | IPR003599:Immunoglobulin subtype | 19 | 2.339901478 | 7.83E-08 | 691 | 30 | 4190 | 3.840328027 | 1.23E-04 |
| SMART | SM00409:IG | 19 | 2.339901478 | 1.60E-06 | 377 | 30 | 1869 | 3.139787798 | 0.001993932 |
| INTERPRO | IPR007110:Immunoglobulin-like | 21 | 2.586206897 | 1.29E-05 | 691 | 46 | 4190 | 2.768199836 | 0.020301505 |

**Annotation Cluster 9** **Enrichment Score: 5.36295761301755**

| Category | Term | Count | % | PValue | List Total | Pop Hits | Pop Total | Fold Enrichment | FDR |
|---|---|---|---|---|---|---|---|---|---|
| KEGG_PATHWAY | dme03010:Ribosome | 26 | 3.201970443 | 1.36E-12 | 161 | 36 | 937 | 4.203243616 | 1.37E-09 |
| GOTERM_CC_FAT | GO:0022626~cytosolic ribosome | 26 | 3.201970443 | 8.76E-12 | 336 | 36 | 1869 | 4.017361111 | 1.21E-08 |
| GOTERM_CC_FAT | GO:0044445~cytosolic part | 27 | 3.325123153 | 8.73E-07 | 336 | 57 | 1869 | 2.634868421 | 0.001205052 |
| SP_PIR_KEYWORDS | ribonucleoprotein | 28 | 3.448275862 | 1.72E-06 | 786 | 66 | 4770 | 2.574600972 | 0.002282836 |
| SP_PIR_KEYWORDS | ribosomal protein | 31 | 3.81773399 | 4.51E-05 | 786 | 89 | 4770 | 2.113817652 | 0.059807775 |
| GOTERM_CC_FAT | GO:0005840~ribosome | 34 | 4.187192118 | 8.61E-05 | 336 | 98 | 1869 | 1.929846939 | 0.118777504 |
| GOTERM_CC_FAT | GO:0033279~ribosomal subunit | 32 | 3.9408867 | 8.78E-05 | 336 | 90 | 1869 | 1.977777778 | 0.121142593 |
| GOTERM_MF_FAT | GO:0003735~structural constituent of ribosome | 32 | 3.9408867 | 1.15E-04 | 536 | 93 | 3078 | 1.975926818 | 0.170384523 |
| GOTERM_BP_FAT | GO:0006412~translation | 39 | 4.802955665 | 0.050115863 | 511 | 163 | 2817 | 1.318994393 | 58.72942407 |
| GOTERM_CC_FAT | GO:0030529~ribonucleoprotein complex | 41 | 5.049261084 | 0.066869989 | 336 | 179 | 1869 | 1.274092179 | 61.53930522 |

**Annotation Cluster 10** **Enrichment Score: 4.591159300959751**

| Category | Term | Count | % | PValue | List Total | Pop Hits | Pop Total | Fold Enrichment | FDR |
|---|---|---|---|---|---|---|---|---|---|
| GOTERM_BP_FAT | GO:0007267~cell-cell signaling | 35 | 4.310344828 | 1.37E-05 | 511 | 93 | 2817 | 2.074679629 | 0.023625556 |
| GOTERM_BP_FAT | GO:0007268~synaptic transmission | 32 | 3.9408867 | 2.68E-05 | 511 | 84 | 2817 | 2.100083869 | 0.046138564 |
| GOTERM_BP_FAT | GO:0019226~transmission of nerve impulse | 32 | 3.9408867 | 4.58E-05 | 511 | 86 | 2817 | 2.051244709 | 0.078769078 |

**Annotation Cluster 11** **Enrichment Score: 4.410252952822596**

| Category | Term | Count | % | PValue | List Total | Pop Hits | Pop Total | Fold Enrichment | FDR |
|---|---|---|---|---|---|---|---|---|---|
| GOTERM_BP_FAT | GO:0046552~photoreceptor cell fate | 16 | 1.97044335 | 2.22E-05 | 511 | 28 | 2817 | 3.150125804 | 0.038285718 |
| | commitment | | | | | | | | |
| GOTERM_BP_FAT | GO:0048663~neuron fate commitment | 16 | 1.97044335 | 2.22E-05 | 511 | 28 | 2817 | 3.150125804 | 0.038285718 |
| GOTERM_BP_FAT | GO:0001752~compound eye photoreceptor fate commitment | 15 | 1.84729064 | 6.80E-05 | 511 | 27 | 2817 | 3.062622309 | 0.116912828 |
| GOTERM_BP_FAT | GO:0042706~eye photoreceptor cell fate commitment | 15 | 1.84729064 | 6.80E-05 | 511 | 27 | 2817 | 3.062622309 | 0.116912828 |

**Annotation Cluster 12** **Enrichment Score: 4.327193195181524**

| Category | Term | Count | % | PValue | List Total | Pop Hits | Pop Total | Fold Enrichment | FDR |
|---|---|---|---|---|---|---|---|---|---|
| GOTERM_BP_FAT | GO:0048609~reproductive process in a multicellular organism | 69 | 8.497536946 | 5.22E-06 | 511 | 229 | 2817 | 1.661037951 | 0.00898789 |
| GOTERM_BP_FAT | GO:0032504~multicellular organism reproduction | 69 | 8.497536946 | 5.22E-06 | 511 | 229 | 2817 | 1.661037951 | 0.00898789 |
| GOTERM_BP_FAT | GO:0007276~gamete generation | 63 | 7.75862069 | 3.94E-05 | 511 | 215 | 2817 | 1.615355209 | 0.067752345 |
| GOTERM_BP_FAT | GO:0019953~sexual reproduction | 63 | 7.75862069 | 5.37E-05 | 511 | 217 | 2817 | 1.600467142 | 0.092370275 |
| GOTERM_BP_FAT | GO:0048610~reproductive cellular process | 51 | 6.280788177 | 8.20E-05 | 511 | 167 | 2817 | 1.683525317 | 0.140974756 |
| GOTERM_BP_FAT | GO:0007292~female gamete generation | 50 | 6.157635468 | 3.19E-04 | 511 | 171 | 2817 | 1.611906479 | 0.547637831 |
| GOTERM_BP_FAT | GO:0048477~oogenesis | 49 | 6.034482759 | 3.40E-04 | 511 | 167 | 2817 | 1.617504717 | 0.583843907 |

**Annotation Cluster 13** **Enrichment Score: 3.5098655719773366**

| Category | Term | Count | % | PValue | List Total | Pop Hits | Pop Total | Fold Enrichment | FDR |
|---|---|---|---|---|---|---|---|---|---|
| GOTERM_BP_FAT | GO:0048666~neuron development | 48 | 5.911330049 | 2.12E-05 | 511 | 147 | 2817 | 1.800071888 | 0.036521547 |
| GOTERM_BP_FAT | GO:0000904~cell morphogenesis involved in differentiation | 43 | 5.295566502 | 4.52E-05 | 511 | 130 | 2817 | 1.823438206 | 0.077726513 |
| GOTERM_BP_FAT | GO:0048667~cell morphogenesis involved in neuron differentiation | 40 | 4.926108374 | 1.35E-04 | 511 | 123 | 2817 | 1.792754522 | 0.23201353 |
| GOTERM_BP_FAT | GO:0048812~neuron projection morphogenesis | 39 | 4.802955665 | 2.06E-04 | 511 | 121 | 2817 | 1.776827158 | 0.354155351 |
| GOTERM_BP_FAT | GO:0031175~neuron projection development | 39 | 4.802955665 | 2.49E-04 | 511 | 122 | 2817 | 1.762263001 | 0.428191192 |
| GOTERM_BP_FAT | GO:0048858~cell projection morphogenesis | 39 | 4.802955665 | 0.001609619 | 511 | 133 | 2817 | 1.616511926 | 2.734806782 |
| GOTERM_BP_FAT | GO:0030030~cell projection organization | 42 | 5.172413793 | 0.002346925 | 511 | 149 | 2817 | 1.553921118 | 3.963852685 |
| GOTERM_BP_FAT | GO:0032990~cell part morphogenesis | 39 | 4.802955665 | 0.003321885 | 511 | 138 | 2817 | 1.557942653 | 5.566603734 |

**S7 – Genes enriched for high constraint**

Below are the top 812 genes enriched for high constraint at 4D sites.

| | | | | | |
|---|---|---|---|---|---|
| **FLYBASE GENE ID** | FBgn0021800 | FBgn0024273 | FBgn0033934 | FBgn0030847 | FBgn0003502 |
| FBgn0037245 | FBgn0022764 | FBgn0029936 | FBgn0046258 | FBgn0010422 | FBgn0017579 |
| FBgn0033029 | FBgn0025712 | FBgn0053995 | FBgn0040571 | FBgn0051637 | FBgn0035283 |
| FBgn0037739 | FBgn0003279 | FBgn0003659 | FBgn0037777 | FBgn0027052 | FBgn0001297 |
| FBgn0028373 | FBgn0032021 | FBgn0033108 | FBgn0033558 | FBgn0000618 | FBgn0039427 |
| FBgn0027550 | FBgn0040636 | FBgn0030870 | FBgn0040318 | FBgn0027565 | FBgn0036844 |
| FBgn0031632 | FBgn0029974 | FBgn0038100 | FBgn0036039 | FBgn0003360 | FBgn0036317 |
| FBgn0014163 | FBgn0016131 | FBgn0004654 | FBgn0000289 | FBgn0003205 | FBgn0020245 |
| FBgn0016061 | FBgn0037313 | FBgn0034636 | FBgn0039237 | FBgn0030260 | FBgn0027546 |
| FBgn0032162 | FBgn0042199 | FBgn0020647 | FBgn0040715 | FBgn0046687 | FBgn0035630 |
| FBgn0031637 | FBgn0039200 | FBgn0004862 | FBgn0035753 | FBgn0031816 | FBgn0026376 |
| FBgn0038763 | FBgn0035866 | FBgn0032297 | FBgn0013433 | FBgn0034421 | FBgn0037647 |
| FBgn0010704 | FBgn0034454 | FBgn0038063 | FBgn0050185 | FBgn0039969 | FBgn0000038 |
| FBgn0017581 | FBgn0052137 | FBgn0004118 | FBgn0008636 | FBgn0037697 | FBgn0026533 |
| FBgn0031126 | FBgn0025806 | FBgn0027600 | FBgn0001134 | FBgn0024232 | FBgn0000581 |
| FBgn0002789 | FBgn0031322 | FBgn0051365 | FBgn0030294 | FBgn0015754 | FBgn0020617 |
| FBgn0011676 | FBgn0031835 | FBgn0038092 | FBgn0017550 | FBgn0024187 | FBgn0039130 |
| FBgn0035873 | FBgn0040823 | FBgn0023423 | FBgn0015919 | FBgn0036044 | FBgn0036428 |
| FBgn0038126 | FBgn0033639 | FBgn0032181 | FBgn0035329 | FBgn0000308 | FBgn0047135 |
| FBgn0014026 | FBgn0026263 | FBgn0037521 | FBgn0036921 | FBgn0033480 | FBgn0039561 |
| FBgn0010894 | FBgn0000183 | FBgn0030361 | FBgn0025574 | FBgn0027657 | FBgn0015521 |
| FBgn0038805 | FBgn0039636 | FBgn0011227 | FBgn0004583 | FBgn0022985 | FBgn0065032 |
| FBgn0035500 | FBgn0035987 | FBgn0010762 | FBgn0035578 | FBgn0027497 | FBgn0036728 |
| FBgn0033174 | FBgn0082582 | FBgn0034861 | FBgn0034460 | FBgn0038619 | FBgn0051221 |
| FBgn0035236 | FBgn0034788 | FBgn0013954 | FBgn0034538 | FBgn0043012 | FBgn0031940 |
| FBgn0003386 | FBgn0014879 | FBgn0038098 | FBgn0035475 | FBgn0029687 | FBgn0022085 |
| FBgn0042112 | FBgn0004636 | FBgn0035400 | FBgn0004873 | FBgn0020238 | FBgn0035495 |
| FBgn0014020 | FBgn0038581 | FBgn0043364 | FBgn0015550 | FBgn0051005 | FBgn0028961 |
| FBgn0033907 | FBgn0037336 | FBgn0039159 | FBgn0035033 | FBgn0015778 | FBgn0033122 |
| FBgn0038515 | FBgn0023179 | FBgn0052485 | FBgn0000150 | FBgn0037814 | FBgn0035528 |
| FBgn0035981 | FBgn0039380 | FBgn0051216 | FBgn0017572 | FBgn0000339 | FBgn0024244 |
| FBgn0036337 | FBgn0000273 | FBgn0032588 | FBgn0026077 | FBgn0003941 | FBgn0042185 |
| FBgn0061361 | FBgn0035936 | FBgn0004572 | FBgn0034886 | FBgn0038589 | FBgn0003527 |
| FBgn0037664 | FBgn0002626 | FBgn0031955 | FBgn0028646 | FBgn0033925 | FBgn0039489 |
| FBgn0050373 | FBgn0035542 | FBgn0010382 | FBgn0035807 | FBgn0022382 | FBgn0033624 |
| FBgn0034091 | FBgn0034245 | FBgn0004363 | FBgn0027779 | FBgn0036364 | FBgn0003312 |
| FBgn0028331 | FBgn0051635 | FBgn0052260 | FBgn0037328 | FBgn0039581 | FBgn0004369 |
| FBgn0027356 | FBgn0034371 | FBgn0031627 | FBgn0001098 | FBgn0003721 | FBgn0036992 |
| FBgn0033985 | FBgn0030893 | FBgn0053100 | FBgn0039816 | FBgn0001319 | FBgn0038294 |
| FBgn0020300 | FBgn0027950 | FBgn0031310 | FBgn0000097 | FBgn0030357 | FBgn0000529 |
| FBgn0050158 | FBgn0042135 | FBgn0010226 | FBgn0051632 | FBgn0000181 | FBgn0010909 |
| FBgn0001078 | FBgn0022343 | FBgn0028871 | FBgn0033236 | FBgn0035335 | FBgn0004552 |
| FBgn0004177 | FBgn0010415 | FBgn0051121 | FBgn0034030 | FBgn0034399 | FBgn0013751 |
| FBgn0032504 | FBgn0011455 | FBgn0003326 | FBgn0020303 | FBgn0004907 | FBgn0036003 |
| FBgn0035281 | FBgn0011481 | FBgn0041087 | FBgn0031589 | FBgn0033309 | FBgn0042138 |
| FBgn0039617 | FBgn0037680 | FBgn0025286 | FBgn0021967 | FBgn0002945 | FBgn0052177 |
| FBgn0036556 | FBgn0004795 | FBgn0027589 | FBgn0030101 | FBgn0029715 | FBgn0031106 |
| FBgn0002973 | FBgn0027951 | FBgn0035085 | FBgn0001942 | FBgn0033544 | FBgn0039223 |
| FBgn0050295 | FBgn0031413 | FBgn0039748 | FBgn0005671 | FBgn0027527 | FBgn0034345 |
| FBgn0037728 | FBgn0000250 | FBgn0010877 | FBgn0014859 | FBgn0032465 | FBgn0035688 |
| FBgn0030850 | FBgn0034802 | FBgn0011211 | FBgn0010397 | FBgn0013973 | FBgn0037614 |
| FBgn0004921 | FBgn0023212 | FBgn0029944 | FBgn0015790 | FBgn0005533 | FBgn0051481 |
| FBgn0051146 | FBgn0038947 | FBgn0002567 | FBgn0015371 | FBgn0005640 | FBgn0011281 |
| FBgn0024734 | FBgn0032629 | FBgn0010269 | FBgn0000577 | FBgn0038654 | FBgn0024963 |
| FBgn0052405 | FBgn0035043 | FBgn0004611 | FBgn0003429 | FBgn0004179 | FBgn0027619 |
| FBgn0004618 | FBgn0004396 | FBgn0031661 | FBgn0037847 | FBgn0033782 | FBgn0010228 |
| FBgn0028622 | FBgn0028406 | FBgn0021895 | FBgn0052240 | FBgn0029708 | FBgn0043458 |
| FBgn0036111 | FBgn0030834 | FBgn0002940 | FBgn0030328 | FBgn0045038 | FBgn0026379 |
| FBgn0010348 | FBgn0032447 | FBgn0002283 | FBgn0034730 | FBgn0038267 | FBgn0010399 |
| FBgn0035844 | FBgn0005777 | FBgn0004919 | FBgn0011217 | FBgn0036851 | FBgn0051176 |
| FBgn0004101 | FBgn0039790 | FBgn0040297 | FBgn0038981 | FBgn0040551 | FBgn0027605 |
| FBgn0034021 | FBgn0028474 | FBgn0036223 | FBgn0011739 | FBgn0035367 | FBgn0010411 |
| FBgn0038683 | FBgn0030257 | FBgn0004889 | FBgn0033486 | FBgn0033452 | FBgn0051660 |
| FBgn0026250 | FBgn0039381 | FBgn0032101 | FBgn0039735 | FBgn0052428 | FBgn0005677 |
| FBgn0030456 | FBgn0032901 | FBgn0035853 | FBgn0012344 | FBgn0025455 | FBgn0034755 |
| FBgn0015324 | FBgn0038659 | FBgn0004892 | FBgn0010078 | FBgn0011278 | FBgn0028516 |
| FBgn0037448 | FBgn0037891 | FBgn0038855 | FBgn0038946 | FBgn0028582 | FBgn0002577 |
| FBgn0035601 | FBgn0011766 | FBgn0001139 | FBgn0000039 | FBgn0020910 | FBgn0034084 |
| FBgn0036436 | FBgn0033784 | FBgn0044323 | FBgn0037539 | FBgn0034408 | FBgn0011589 |
| FBgn0003275 | FBgn0038140 | FBgn0029067 | FBgn0039116 | FBgn0040384 | FBgn0004387 |
| FBgn0035982 | FBgn0021979 | FBgn0015299 | FBgn0040286 | FBgn0000099 | FBgn0038839 |
| FBgn0013726 | FBgn0033209 | FBgn0000108 | FBgn0029128 | FBgn0003411 | FBgn0029975 |
| FBgn0035323 | FBgn0030791 | FBgn0028342 | FBgn0029896 | FBgn0035558 | FBgn0050419 |
| FBgn0000157 | FBgn0031799 | FBgn0032305 | FBgn0030089 | FBgn0038494 | FBgn0024753 |
| FBgn0037747 | FBgn0016700 | FBgn0029911 | FBgn0000591 | FBgn0035533 | FBgn0011648 |
| FBgn0028789 | FBgn0036781 | FBgn0019662 | FBgn0062413 | FBgn0037358 | FBgn0003425 |
| FBgn0030873 | FBgn0000061 | FBgn0033668 | FBgn0050446 | FBgn0035792 | FBgn0050147 |
| FBgn0039225 | FBgn0001235 | FBgn0031866 | FBgn0030234 | FBgn0015591 | FBgn0011656 |
| FBgn0036685 | FBgn0033736 | FBgn0041184 | FBgn0013325 | FBgn0016726 | FBgn0034138 |
| FBgn0031359 | FBgn0051191 | FBgn0025633 | FBgn0030183 | FBgn0063649 | FBgn0027607 |
| FBgn0000635 | FBgn0039266 | FBgn0053505 | FBgn0016797 | FBgn0014001 | FBgn0039260 |
| FBgn0016691 | FBgn0034501 | FBgn0036126 | FBgn0036974 | FBgn0037770 | FBgn0011581 |
| FBgn0038964 | FBgn0024944 | FBgn0038877 | FBgn0032223 | FBgn0000564 | FBgn0001122 |
| FBgn0040512 | FBgn0052672 | FBgn0000115 | FBgn0015904 | FBgn0003896 | FBgn0034763 |
| FBgn0052843 | FBgn0001085 | FBgn0029768 | FBgn0051772 | FBgn0052057 | FBgn0031945 |
| FBgn0038065 | FBgn0035505 | FBgn0032499 | FBgn0083228 | FBgn0011224 | FBgn0010280 |
| FBgn0031692 | FBgn0051641 | FBgn0015799 | FBgn0038787 | FBgn0002593 | FBgn0052100 |
| FBgn0005626 | FBgn0011272 | FBgn0011661 | FBgn0041789 | FBgn0032731 | FBgn0040726 |
| FBgn0033869 | FBgn0003410 | FBgn0040985 | FBgn0003720 | FBgn0020912 | FBgn0004169 |
| FBgn0032378 | FBgn0039448 | FBgn0029943 | FBgn0037408 | FBgn0036382 | FBgn0052183 |
| FBgn0005633 | FBgn0013467 | FBgn0038931 | FBgn0031602 | FBgn0020618 | FBgn0020497 |
| FBgn0030052 | FBgn0028408 | FBgn0013799 | FBgn0020620 | FBgn0033317 | FBgn0032642 |
| FBgn0027492 | FBgn0035285 | FBgn0031646 | FBgn0046763 | FBgn0036761 | FBgn0002609 |
| FBgn0000625 | FBgn0031005 | FBgn0025800 | FBgn0003513 | FBgn0003975 | FBgn0031257 |
| FBgn0051145 | FBgn0031150 | FBgn0000395 | FBgn0004777 | FBgn0033931 | FBgn0000253 |
| FBgn0051163 | FBgn0002561 | FBgn0015320 | FBgn0000037 | FBgn0004551 | FBgn0001123 |
| FBgn0028969 | FBgn0003997 | FBgn0036134 | FBgn0026086 | FBgn0022268 | FBgn0033783 |
| FBgn0004646 | FBgn0032428 | FBgn0025743 | FBgn0038592 | FBgn0010105 | FBgn0004242 |
| FBgn0039359 | FBgn0035186 | FBgn0029764 | FBgn0030786 | FBgn0039831 | FBgn0053126 |
| FBgn0052105 | FBgn0002772 | FBgn0031836 | FBgn0034903 | FBgn0000024 | FBgn0039830 |
| FBgn0053094 | FBgn0000166 | FBgn0038145 | FBgn0000152 | FBgn0028563 | FBgn0000490 |
| FBgn0037721 | FBgn0029092 | FBgn0020309 | FBgn0015561 | FBgn0001994 | FBgn0052299 |
| FBgn0052698 | FBgn0051869 | FBgn0004436 | FBgn0011745 | FBgn0037551 | FBgn0033699 |
| FBgn0040079 | FBgn0016926 | FBgn0030364 | FBgn0031950 | FBgn0045064 | FBgn0039705 |
| FBgn0051361 | FBgn0002922 | FBgn0037972 | FBgn0033912 | FBgn0036341 | FBgn0033915 |
| FBgn0032856 | FBgn0031603 | FBgn0030680 | FBgn0003744 | FBgn0038118 | FBgn0030797 |
| FBgn0034504 | FBgn0029737 | FBgn0011297 | FBgn0026438 | FBgn0051100 | FBgn0019936 |
| FBgn0000258 | FBgn0043900 | FBgn0034539 | FBgn0040827 | FBgn0030672 | FBgn0017549 |
| FBgn0038658 | FBgn0004882 | FBgn0027581 | FBgn0032840 | FBgn0004638 | FBgn0000036 |
| FBgn0010516 | FBgn0039523 | FBgn0037351 | FBgn0011277 | FBgn0035429 | FBgn0035436 |
| FBgn0052372 | FBgn0016078 | FBgn0034650 | FBgn0039678 | FBgn0033961 | FBgn0031186 |
| FBgn0030976 | FBgn0032719 | FBgn0036391 | FBgn0003984 | FBgn0036032 | FBgn0014010 |
| FBgn0010114 | FBgn0024188 | FBgn0031850 | FBgn0002773 | FBgn0037424 | FBgn0039154 |
| FBgn0031826 | FBgn0011640 | FBgn0003612 | FBgn0014454 | FBgn0035060 | FBgn0043070 |
| FBgn0034013 | FBgn0017397 | FBgn0036801 | FBgn0026316 | FBgn0037698 | FBgn0035142 |
| FBgn0051140 | FBgn0003944 | FBgn0033971 | FBgn0041092 | FBgn0035016 | FBgn0052103 |
| FBgn0034674 | FBgn0039595 | FBgn0008646 | FBgn0004666 | FBgn0023170 | FBgn0036661 |
| FBgn0025879 | FBgn0025391 | FBgn0037926 | FBgn0022238 | FBgn0015721 | FBgn0037429 |
| FBgn0051291 | FBgn0004395 | FBgn0035600 | FBgn0030603 | FBgn0040600 | FBgn0052056 |
| FBgn0034889 | FBgn0034743 | FBgn0020235 | FBgn0026389 | FBgn0030038 | FBgn0020767 |
| FBgn0001138 | FBgn0035945 | FBgn0027556 | FBgn0019948 | FBgn0000575 | FBgn0010424 |
| FBgn0015806 | FBgn0037430 | FBgn0000633 | FBgn0005659 | FBgn0002629 | FBgn0039272 |
| FBgn0030529 | FBgn0038498 | FBgn0038389 | FBgn0028734 | FBgn0034644 | FBgn0051337 |
| FBgn0032083 | FBgn0003274 | FBgn0033129 | FBgn0003204 | FBgn0034602 | FBgn0031971 |
| FBgn0039151 | FBgn0026084 | FBgn0040376 | FBgn0039454 | FBgn0032666 | FBgn0032261 |
| FBgn0003319 | FBgn0035675 | FBgn0034645 | FBgn0038282 | FBgn0004841 | FBgn0036380 |
| FBgn0033726 | FBgn0017551 | FBgn0000171 | FBgn0035816 | FBgn0001197 | FBgn0000409 |
| FBgn0050271 | FBgn0029508 | FBgn0039844 | FBgn0051708 | FBgn0011259 | FBgn0004514 |
| FBgn0004908 | FBgn0034585 | FBgn0027932 | FBgn0016687 | FBgn0020496 | FBgn0020372 |
| FBgn0034570 | FBgn0038881 | FBgn0033551 | FBgn0033728 | FBgn0035954 | FBgn0037421 |
| FBgn0025463 | FBgn0034946 | FBgn0023535 | FBgn0037445 | FBgn0036967 | FBgn0045759 |
| FBgn0003267 | FBgn0036257 | FBgn0026597 | FBgn0032633 | FBgn0039584 | FBgn0011701 |
| FBgn0020445 | FBgn0013750 | FBgn0003475 | FBgn0029894 | FBgn0038043 | FBgn0027590 |
| FBgn0061198 | FBgn0000256 | FBgn0031090 | FBgn0036583 | FBgn0029152 | |
| FBgn0028420 | FBgn0026753 | FBgn0023528 | FBgn0050118 | FBgn0030766 | |
| FBgn0004868 | FBgn0037374 | FBgn0037262 | FBgn0030790 | FBgn0035517 | |

**S8 – Phylogenetic tree and parameters of 4D sites**

Below is the ascertained tree for 4D sites in *D. melanogater*, representing the average rate of evolution along each branch for 4D sites:

(((((dm3:0.0570839230,(droSim1:0.0191559629,droSec1:0.0228909640):0.0245005060):0.0521408013,(droYak2:0.0880756403,droEre2:0.0765664754):0.0327248791):0.3298133685,droAna3:0.4828717943):0.1740811596,(dp4:0.0119588627,droPer1:0.0118166618):0.4002579069):0.1152779250,droWil1:0.6852845130,((droVir3:0.2436546904,droMoj3:0.3505049405):0.0822301249,droGri2:0.3355523333):0.2483077527);

dm3 – *D. melanogaster*, droSim1 – *D. simulans*, droSec1 – *D. sechellia,* droYak2 – *D. yakuba*, droEre2 – *D. erecta*, droAna3 – *D. ananassae*, dp4 – *D. pseudoobscura*, droPer1 – *D. persimilis*, droWil1 – *D. willistoni*, droVir3 – *D. virilis*, droMoj3 – *D. mojavensis*, droGri2 – *D. grimshawi*

transition/transversion rate ratio, k = 1.865
$\pi_A$ = the probability of being in state A
$\pi_A = 0.19985$
$\pi_C = 0.38418$
$\pi_G = 0.23557$
$\pi_T = 0.18040$

GERP uses a different parameterization of the transition-transversion ratiofrom that used by PhyML. (citations) Using the HKY85 model (citation), PhyML's k is a rate modifier for transitions (i.e. $r_{A\text{->}G} = k*\pi_G$), while GERP's R is the ratio in overall frequency of change (i.e. $f_{A\text{->}G} = \pi_A*r_{A\text{->}G}$ – the probability of being in A and going to G) between transitions and transversions (i.e. $R = f_{transiution}/f_{transversion}$).

The two parameterizations, R (GERP) and k (PhyML) are related as follows:

$$R = k*(\pi_A*\pi_G + \pi_C*\pi_T)/(\pi_X*\pi_Y)$$
$$\pi_X = \pi_A + \pi_G$$
$$\pi_Y = \pi_C + \pi_T$$

$\therefore$ R = 0.8830